\documentclass[aps,prd,showpacs,amssymb,floatfix,nofootinbib]{revtex4}
\usepackage{graphicx,amsmath,subfigure}

\newcommand{\eeq}{\end{equation}}
\newcommand{\bea}{\begin{eqnarray}}
\newcommand{\rmd}{{\rm d}}
\def\ltsima{$\; \buildrel < \over \sim \;$}
\def\simlt{\lower.5ex\hbox{\ltsima}}
\def\gtsima{$\; \buildrel > \over \sim \;$}
\def\simgt{\lower.5ex\hbox{\gtsima}}

\def\lesssim{\mathrel{\hbox{\rlap{\hbox{\lower4pt\hbox{$\sim$}}}\hbox{$<$}}}}
\def\gtrsim{\mathrel{\hbox{\rlap{\hbox{\lower4pt\hbox{$\sim$}}}\hbox{$>$}}}}
\def\alt{\mathrel{\hbox{\rlap{\hbox{\lower4pt\hbox{$\sim$}}}\hbox{$<$}}}}
\def\agt{\mathrel{\hbox{\rlap{\hbox{\lower4pt\hbox{$\sim$}}}\hbox{$>$}}}}

\def\gta{\ifmmode {\mathbin{\lower 3pt\hbox   
    {$\,\rlap{\raise 5pt\hbox{$\char'076$}}\mathchar"7218\,$}}}
    \else {${\mathbin{\lower 3pt\hbox
    {$\rlap{\raise 5pt\hbox{$\char'076$}}\mathchar"7218\,$}}}
    $}\fi}
\def\lta{\ifmmode {\,\mathbin{\lower 3pt\hbox   
    {$\,\rlap{\raise 5pt\hbox{$\char'074$}}\mathchar"7218\,$}}}
    \else {${\mathbin{\lower 3pt\hbox
    {$\rlap{\raise 5pt\hbox{$\char'074$}}\mathchar"7218\,$}}}
    $}\fi}

\begin{document}
\title{ Intermediate--mass--ratio--inspirals in the Einstein Telescope: \\ I. Signal--to--noise ratio calculations.}
\author{E. A. Huerta and Jonathan R. Gair}

\affiliation{Institute of Astronomy, Madingley Road, CB3 0HA Cambridge, UK}

\email{eah41@ast.cam.ac.uk}
\email{jgair@ast.cam.ac.uk}


\date{\today}

\begin{abstract}        
The Einstein Telescope (ET) is a proposed third generation ground--based interferometric gravitational wave  detector, for which the target is a sensitivity that is a factor of ten better than Advanced LIGO and a frequency range that extends down to $\sim 1$Hz. Such a third generation interferometer will provide opportunities to test Einstein's theory of relativity in the strong field and will realize precision gravitational wave astronomy with a thousandfold increase in the expected number of events over the advanced ground-based detectors. A design study for ET is currently underway, so it is timely to assess the science that could be done with such an instrument. This paper is the first in a series that will carry out a detailed study of intermediate--mass--ratio inspirals (IMRIs) for ET. In the context of ET, an IMRI is the inspiral of a neutron star or stellar-mass black hole  into an intermediate mass black hole (IMBH). In this paper we focus on the development of IMRI waveform models for circular and equatorial inspirals. We consider two approximations for the waveforms, which both incorporate the inspiral, merger and ringdown phases in a consistent way. One approximation,  valid for IMBHs of arbitrary spin, uses the transition model of Ori and Thorne~\cite{amos} to describe the merger and this is then matched smoothly onto a ringdown waveform. The second approximation uses the Effective One Body (EOB) approach to model the merger phase of the waveform and is valid for non-spinning IMBHs. In this paper, we use both waveform models to compute signal--to--noise ratios (SNRs) for IMRI sources detectable by ET.  At a redshift of $z=1$, we find typical SNRs for IMRI systems with masses $1.4M_\odot$+$100M_{\odot}$, $10M_\odot$+$100M_{\odot}$, $1.4M_\odot$+$500M_{\odot}$ and $10M_\odot$+$500M_{\odot}$ of $\sim 10$--$25$, $\sim40$--$80$, $\sim3$--$15$ and $\sim 10$--$60$ respectively. We also find that the two models make predictions for non--spinning inspirals that are consistent to about ten percent.
\end{abstract}


\maketitle

\section{Introduction}    

The Einstein Telescope (ET) is a proposed third generation ground--based gravitational wave (GW) detector for which the target is a sensitivity ten times better than that of the advanced detectors~\cite{Freise:2009,Hild:2009}. By siting the instrument underground, it is also hoped that the range of frequencies to which the detector is sensitive can be extended into the 1Hz to 10Hz range, while also maintaining high frequency sensitivity up to 10 kHz. A design study for ET is currently underway, which is exploring the design, cost, site selection, etc., plus the potential scientific impact of such an instrument, with a view to maximizing the scientific output within a reasonable budget. If it is realized, ET will open up the possibility to study a wide variety of sources and address outstanding problems in fundamental physics, cosmology and astrophysics, e.g., studying compact objects (COs) and general relativistic instabilities, solving the enigma of gamma--ray--bursts and shedding light on their different classes, understanding the mass--spectrum of compact stars and their populations, and measuring cosmological parameters using GW sources as standard sirens~\cite{punturoet}. 

The Advanced LIGO and Virgo detectors should achieve sensitivity in the low--frequency regime down to the seismic limit at roughly 10 Hz. For these detectors, the expected isotropic detection horizon (the angle--averaged distance to which a binary can be measured) for Advanced Virgo will be at $\sim150$ Mpc (z = 0.035) for neutron star--neutron star  (NS--NS) binaries and $\sim310$ Mpc (z = 0.07) for neutron star--black hole (NS--BH) binaries \cite{noise1}. For Advanced LIGO, NS--NS signals will be in band for $\sim17$ minutes and have a single detector detection horizon of $\sim200$ Mpc (z = 0.045); NS--BH signals will be in band for $\sim4$ minutes, and have a detection horizon of $\sim420$ Mpc (z = 0.09). In contrast, if ET's seismic wall is pushed down to 5 Hz NS--NS signals will be in band for $\sim2$ hours, and  NS--BH signals for $\sim25$ minutes. If the seismic wall is further reduced down to 3 Hz/1 Hz, then NS--NS signals will be in band for $\sim7$hours/$\sim5$ days, and NS--BH signals will be in band for $\sim2$ hours/$\sim1$ day, respectively. The Einstein Telescope detection horizon is expected to be at  \(z \simeq 1\) for NS--NS signals, and at \(z \simeq 2\) for NS--BH signals \cite{noise1}.

The preceding paragraph served to illustrate that ET will be able to do the same type of science as the advanced detectors, but better. However, there is also some science which only ET will be able to do. The $1$--$10$Hz frequency band lies below the range of current and advanced ground--based interferometers and above the range of the proposed space-based detector, LISA~\cite{SRD}. If ET achieves good sensitivity in this range, it will open up a new gravitational wave window. GWs in this range would be generated by sources with intermediate mass, i.e., from hundreds to a few thousand solar masses. Such intermediate--mass black holes (IMBHs) might be primordial, i.e., form in the early Universe as seeds from which BHs in galaxies subsequently grow, or they may form in the centre of dense globular clusters through runaway stellar collisions (see~\cite{etgair} and references therein for descriptions of these two scenarios). Observational evidence for the existence of IMBHs has accumulated over the last decade. This evidence is of  two different types. First,  ultraluminous X--ray sources (ULXs) have been observed that are not associated with active galactic nuclei and yet have fluxes many times the angle--averaged flux of a \(M < 20 M_\odot\) BH accreting at the Eddington limit. Second, in several clusters, e.g., the globular clusters M15 and G1, the stellar kinematics shows evidence for an excess of dark mass in the centre. In M15, there is also evidence for rotation in the core at a speed that is comparable to the central stellar velocity dispersion. One possible explanation for this rotation would be the transferal of angular momentum from a black hole binary to stars in the core through three body encounters. A binary comprising a \(\sim 20 M_{\odot}\) black hole orbiting a \(10^{3} M_{\odot}\) object at a semimajor axis of \(10^{-3}\) pc would have enough angular momentum to account for the observed rotation (see~\cite{evidence} for a discussion of this and other observational evidence for the existence of IMBHs).

The observational evidence will further improve with future X-ray and optical observations, but a robust mass determination of an IMBH candidate will be needed for a solid identification. This may be possible using radial velocity measurements of companions to ULXs in binaries, but this technique only yields a lower bound on the mass due to the unknown inclination of the system, and the companions are typically very faint. If IMBHs do exist, they will merge with other IMBHs and smaller COs, and these mergers will generate gravitational waves in the ET band. GW observations will provide very accurate mass measurements, and so the first convincing proof of existence for IMBHs may come from these observations~\cite{evidence}. 

In this paper we will be concentrating on GWs from the mergers of IMBHs with lower mass COs, which might be either NSs or stellar-mass BHs. These sources are termed intermediate-mass-ratio inspirals (IMRIs). The dominant mechanism that leads to the formation of IMRIs is three-body hardening of an IMBH-CO binary in a core-collapsed globular cluster. 

Numerical simulations of globular clusters suggest that  a fraction \(\sim10^{-6}\) to \(\sim 10^{-4}\) of the \(\sim 10^6\) initial stars that form a globular cluster will become stellar--mass BHs via normal stellar evolution \cite{nat}.  Assuming a globular cluster  with relaxation time of 1 Gyr \cite{oleary}, all these BHs should have formed within \(\sim 10\) Myr, with the most massive BHs forming at around \(\sim 3\) Myr \cite{schaller}. These black holes should be more centrally concentrated than main sequence (MS) stars since there will be significant mass segregation of their higher mass progenitors \cite{freitag}, there will be preferential formation of stars near the cluster centre \cite{murray}, and because BH birth kicks are not expected to displace BHs into the cluster halo \cite{white}. Even if we assume that the BHs were distributed throughout the cluster, mass segregation should be able to assemble a subcluster of BHs near the centre after at most \(\sim 100\) Myr. During mass segregation, BH--MS binaries will undergo three body and four body interactions that will replace the MS star by a heavier BH. Simulations~\cite{oleary,tagushi} suggest that, whether formed through successive BH mergers or stellar collisions, it is more likely to find an IMBH in a cluster with a dense core, so that IMBHs could exist in some tens of percent of current globular clusters. In this type of clusters, the first formation of an IMBH with mass \(\sim 100 M_{\odot}\) could be\(\sim 10\) Myr after the subcluster of BHs is formed. IMBHs formed in this way have a negligible cross section for direct collisions/plunges with other objects. However, they will readily swap into a binary, as the heaviest object in the cluster, and such binaries can be driven to plunge through three body hardening or the secular Kozai resonance. IMBH-CO binaries will eventually be driven to merger through gravitational wave emission, and the GWs generated during these final phases of evolution could be detected by ground--based laser interferometers \cite{etgair}. It has been estimated that in such environments the IMRI event rate for ET could be as high as a few hundred per year~\cite{etgair}.

We will describe two alternative models for the gravitational waveform generated during an IMRI,  and then we will use them to estimate the signal-to-noise ratio (SNR) of events detected by ET. The low-frequency cut-off in the ET sensitivity will mean that only a short section of the inspiral phase will be observed for a typical ET IMRI event. A significant fraction of the total SNR will therefore come from the merger and ringdown and so it is important to include them in the waveform model. In both models, we will use the ``numerical kludge''~\cite{kludge} approach to describe the inspiral phase. This model was developed for extreme-mass-ratio systems for LISA and is based on the adiabatic inspiral of a test-mass through a sequence of geodesic orbits, with the sequence being determined by post-Newtonian energy and angular momentum fluxes, augmented with fits to Teukolsky fluxes and post-Newtonian conservative corrections~\cite{cons}. The two models we consider here differ in the treatment of the merger. In the first model, which is valid for inspirals into IMBHs of arbitrary spin, we model the transition from adiabatic inspiral to plunge using the scheme provided by Ori and Thorne~\cite{amos}. In the second model, which we use to cross--check our results in the non-spinning limit, we model the merger and plunge using the Effective One Body (EOB) approach.

In our SNR calculations, we will assume that a network of ET-like detectors exists, rather than just a single detector. This will be necessary if we are to estimate the extrinsic parameters of IMRI events to any precision. The duration of an IMRI event will be between a few seconds (e.g., a $10M_{\odot}$ plus $500M_{\odot}$ system) and a few minutes (e.g., a $1.4M_{\odot}$ plus $100M_{\odot}$ system). Over such a short timescale, the event is effectively a burst in the detector and so a single ET cannot determine the sky position of the source. While we do not consider parameter estimation in this paper, we assume the existence of a network since some of the science will require it. Paper II in this series will explore parameter estimation accuracies for the same types of ET network.

This paper is organized as follows. In Section~\ref{s2.1} we describe the assumptions we have adopted for the detector and detector network, and describe the binary systems that will be used in the analysis. In Section~\ref{s2.2} we summarise the ``numerical kludge'' model that we shall use to model the inspiral phase. In Section~\ref{s2b} we summarize the transition to plunge scheme introduced by Ori and Thorne~\cite{amos}. In Section~\ref{s2c} we describe the ringdown (RD) radiation and explain how the merger waveforms can be matched onto the RD. In Section~\ref{s2.3} we introduce the aspects of the EOB model relevant for our analysis. In Section~\ref{s2.5} we describe our implementation of the ET response function and describe the dynamics of twelve sample binary systems. We describe the orbital phase evolution up to the light ring (LR), i.e., the inner--most unstable circular orbit for massless particles, and present gravitational waveforms including the final inspiral, merger and RD. This section also includes a comparison between the two models for inspirals into non-spinning IMBHs. In Section~\ref{s2.6} we will present ET SNRs for the sample binary systems and Section~\ref{s6} contains our conclusions.

\section{Assumptions}
\label{s2.1}
\subsection{Einstein Telescope Design}
\label{ETdes}

In this sequence of papers, we take the response of a ``single ET'' to be that of two right--angle interferometers, coplanar and colocated, but rotated \(45^{\circ}\) with respect to each other. Assuming uncorrelated noise, this set--up is equivalent to the currently favoured ET design, i.e., a triangular configuration, with three 10 km long arms, and containing three independent detectors with \(60^{\circ}\) opening angles. This design has the capability to measure polarization at a single site, has lower infrastructure costs, and its sensitivity is a factor \(\sqrt{3}/2\sqrt2 \approx 1.06\) higher than the two right--angle interferometer configuration \cite{etgair}. We shall ignore this factor since it is small compared to other uncertainties in the design.

\begin{figure*}[!htp]
\centerline{
\includegraphics[width=0.5\textwidth, angle=0,clip]{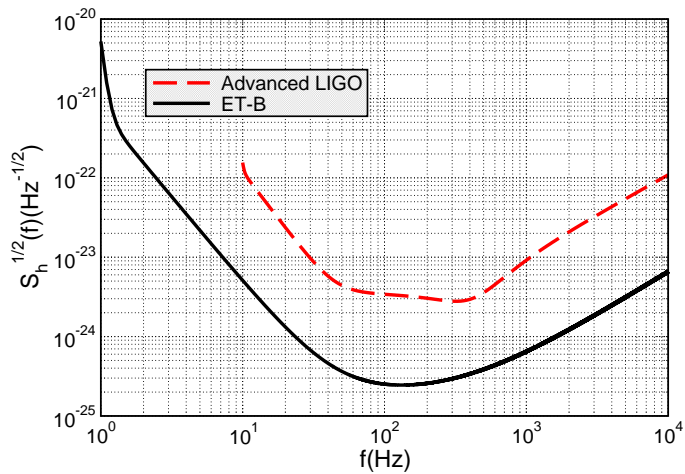}
}
\caption{ Sensitivity curve for the Einstein Telescope, as described in the text. The Advanced LIGO noise curve is also shown for reference.}
\label{figura0}
\end{figure*}

In our studies, we will consider the ``ET B'' sensitivity curve, which at the current stage of the design study is the official sensitivity curve for ET \cite{etb}.  The corresponding amplitude spectrum is shown in Figure~\ref{figura0}.  

An analytic fit to the ET--B spectral density is given by~\cite{etb}
\begin{equation}
S^{1/2}_h(f) = \left\{ \begin{array}{ll}
S^{1/2}_0\left[a_{1} x^{b_1} + a_{2} x^{b_2} + a_{3} x^{b_3} +a_{4} x^{b_4} \right] & \textrm{if  $f \ge f_s$},\\
\infty &  \textrm{if  $f < f_s$},
\end{array} \right .
\label{sensitivity}
\end{equation}
\noindent where \(x=f/f_0\), \(f\) stands for the frequency, \(f_0 = 100 \rm Hz\), \(S_0=10^{-50}{\textrm{Hz} }^{-1}\),  and \(f_s\) is a low frequency cut--off that can be varied, and  below which the sensitivity curve can be considered infinite for practical purposes. The various coefficients take the values
\begin{eqnarray}
a_1 &=&2.39\times10^{-27},\qquad\, b_1=-15.64,\nonumber\\
a_2 &=&0.349, \,~~~~\qquad\qquad b_2=-2.145,\nonumber\\
a_3 &=&1.76, ~~~\qquad\qquad\quad b_3=-0.12,\nonumber\\
a_4 &=&0.409, ~~\qquad\qquad\quad b_4=1.10.
\label{constants_sqrtpsd}
\end{eqnarray}

\noindent Although it is hoped that ET will have sensitivity down to $1$Hz, it is is not yet clear whether this will be achievable, and $3$Hz might be more realistic. To be conservative, for the results described in this paper, we used a cut-off at $5$Hz and started the inspiral evolution at the point when the GWs emitted by the IMRI swept through a frequency of 5Hz.

As mentioned in the introduction, parameter estimation will require the existence of more than one, well-separated, detector. In this paper we will assume the existence of a detector network comprising three detectors sited at the current geographic locations of Virgo, Perth (Australia) and LIGO Livingston. We shall consider five configurations, C1-C5, for a few sample systems. These configurations are C1: one ET at the geographic location of Virgo; C2: as configuration C1 plus a right--angle detector at the location of LIGO Livingston;  C3: as configuration C1 plus another ET at the location of LIGO Livingston;  C4: as configuration C2 plus another right--angle detector in Perth; and C5: as configuration C3 plus another ET in Perth. 

In both this paper, and the next in the series, we shall first quote results for the most optimistic configuration C5, i.e., a network of three detectors each with the sensitivity of a single ET, and then we will go on to compare these with more modest configurations. The aim of this presentation is to exhibit the best performance achievable with a third generation ground--based detector network, but also to set the appropriate framework for the second paper, in which the assumption of a detector network will be required for the parameter estimation studies we will carry out. This presentation also follows the philosophy elsewhere in the literature, of first exploring the best performance achievable by detector networks, both in terms of detection and parameter estimation, for binary inspirals and burst--like events \cite{fairhurst,ares,ajith}.  A 3 ET detector network is extremely optimistic, but we will show that more modest configurations will still produce fairly competitive results. 

\subsection{Sample IMRI systems}
We will present results for twelve different binary systems. We take four combinations for the component masses, namely $1.4M_{\odot} + 100M_{\odot}$,  $1.4M_{\odot} + 500M_{\odot}$, $10M_{\odot}+100M_{\odot}$ and $10M_{\odot}+500M_{\odot}$, and three different values for the spin of the central IMBH, \(q=0,\,0.3,\,0.9\). The twelve sample binaries we examine are all possible combinations of these masses with these spin parameters.

\section{IMRI waveform modelling - Inspiral phase}
For comparable mass binary systems, the early inspiral phase is well modelled by post--Newtonian (PN) theory and the final few cycles can now be computed accurately using numerical relativity \cite{rev}. By contrast, extreme-mass-ratio inspiral (EMRI) systems, in which the mass ratio, \(\eta = m_1 m_2 /(m_1+m_2)^2\), is of the order of $10^{-5}$, emit thousands of cycles in a regime where the velocity is a significant fraction of the speed of light. PN theory therefore does not apply, while numerical relativity cannot be used due to the large number of orbits that must be modelled. However, EMRIs can be accurately modelled using black hole perturbation theory, treating the mass ratio as a small expansion parameter. IMRIs lie somewhere between these two regimes, with mass ratios at which none of the preceding techniques have been tested. Accurate IMRI waveforms are therefore not known at present, and so we will construct our models using the best of what is currently available. We will take the inspiral model from the EMRI limit, but augment it with higher-order-in-mass-ratio PN corrections. We will consider two models for the merger and ringdown, and cross-check their predictions in order to improve our confidence in the results. While the resulting models may not represent the exact waveform of true IMRI sources, they will capture the main features of the signals and so should make reliable predictions for the SNRs and parameter accuracies that ET observations will achieve.

\label{s2.2}
We model the inspiral phase evolution, for both non-spinning and spinning IMBHs, using the ``numerical kludge'' waveform model described in~\cite{cons}. This scheme was developed for the modelling of EMRI systems for LISA and has various nice features: a) the waveforms have been checked against more accurate, Teukolsky--based, waveforms for test-particles on geodesic orbits  and the overlap exceeds \(0.95\) over a large portion of the parameter space \cite{kludge}; b) they are computationally inexpensive; c) conservative self--force corrections to this model have been derived~\cite{cons} for Kerr circular equatorial orbits at 2PN order. This model is not complete as conservative corrections are not yet known for generic orbits, the phase space trajectories are approximate, although they have been matched to Teukolsky based evolutions, and the waveform is constructed from the trajectory using a flat--spacetime wave--emission formula. Despite these various approximations, the numerical kludge waveforms should capture the main features of the inspiral waveform accurately.

During the inspiral phase, radiation--reaction drives the motion of the CO. On short timescales, the small object follows an approximately geodesic orbit in the spacetime of the larger body. Over longer timescales, radiation--reaction causes the orbit to evolve adiabatically. This evolution can be characterized by changes in the geodesic orbital elements, namely the energy, \(E\), angular momentum, \(L_{\mathrm{z}}\), and the Carter constant, \(Q\). In this analysis, we restrict our attention to circular-equatorial orbits, for which the energy and angular momentum take the form  \cite{chandra} 
\begin{eqnarray}
\tilde{E}&=&\frac{E}{\mu} = \frac{1 - 2\left(M/p\right) \pm \left(a/M\right)\left(M/p\right)^{3/2}}{\sqrt{1 - 3\left(M/p\right) \pm 2\left(a/M\right)\left(M/p\right)^{3/2}}}, \nonumber \\   \tilde{L_{\mathrm{z}}}&=&\frac{L_{\mathrm{z}}}{\mu M} = \pm \left(\frac{p}{M}\right)^{1/2}\frac{1  \mp 2\left(a/M\right)\left(M/p\right)^{3/2} + \left(a/M\right)^{2}\left(M/p\right)^{2 }}{\sqrt{1 - 3\left(M/p\right) \pm 2\left(a/M\right)\left(M/p\right)^{3/2}}}, \label{1.1}
\end{eqnarray}
\noindent where \(p\) is the Boyer-Lindquist radius of the orbit, \(a\) is the spin parameter of the central black hole and the upper (lower) sign in Eq. \eqref{1.1} refers to prograde (retrograde) orbits. The Carter constant, $Q$, is zero for equatorial orbits.

The orbital evolution of the CO is obtained by equating the rate of loss of energy  \(E\) and angular momentum \(L_{\mathrm{z}}\) with the corresponding fluxes carried to infinity by the GWs, namely \(\dot{E}\) and \(\dot{L_{\mathrm{z}}}\). These fluxes of energy and angular momentum must satisfy a consistency relation to ensure that circular orbits remain circular under radiation reaction \cite{ori}, i.e.,
\begin{equation}
\dot E(p) = \pm \frac{\sqrt{M}}{p^{3/2} \pm a \sqrt{M}} \dot L_{\mathrm{z}}(p) = \Omega(p) \dot L_{\mathrm{z}}(p),
\label{1.2}
\end{equation}
\noindent where \( \mathrm{d} \phi / \mathrm{d} t = \Omega(p)\), is the azimuthal velocity of the orbit. The inspiral trajectory is given by 
\begin{equation}
\label{1.3}
\dot p= \frac{\mathrm{d} p}{\mathrm{d} E}\dot E= \frac{\mathrm{d} p}{\mathrm{d} L_z}\dot L_z,
\end{equation}
\noindent since the energy and angular momentum fluxes are related to each other by  \eqref{1.2}, so we can use either the energy or the angular momentum flux to evolve the orbit. We choose to evolve \(L_z\), using a PN expression augmented by fits to more accurate fluxes computed using the Teukolsky equation. The expression for \(\dot L_{\mathrm{z}}\) takes the following form for circular equatorial orbits \cite{improved}
\begin{eqnarray}
\dot{L}_z&=& -\frac{32}{5} \frac{\mu^2}{M} \left(\frac{M}{p}\right)^{7/2} 
\Bigg\{ 1 -\frac{61}{12}q\left(\frac{M}{p}\right)^{3/2} - \frac{1247}{336}\left(\frac{M}{p}\right)  +
4\pi \left(\frac{M}{p}\right)^{3/2}   \nonumber \\ &&  - \frac{44711}{9072}\left(\frac{M}{p}\right)^2  
  + \frac{33}{16}\,q^2\left(\frac{M}{p}\right)^2  + \textrm{high order Teukolsky fits}\Bigg\} ,\
\label{new_Ldot}
\end{eqnarray}
\noindent in which we have introduced \(q=a/M\) to denote the dimensionless black hole spin. The last ingredient to include in the inspiral phase evolution is the contribution of the conservative piece of the self--force (SF) to the evolution of the orbit. The conservative piece of the SF affects the frequency of an orbit at a given radius but does not lead to orbital evolution. The frequency shift accumulates over time, which affects the phasing of the waveform. We include this effect by modifying the \(\phi\) angular evolution as follows 
\begin{equation}
\frac{\mathrm{d}\phi}{\mathrm{d}t} = \left(\frac{\mathrm{d}\phi}{\mathrm{d}t}\right)_{\mathrm{geo}}\bigg(1+  \delta \Omega \bigg),
\label{1.4}
\end{equation}
\noindent i.e., we multiply the phase derivative for a Kerr geodesic, labeled by the subscript ``geo'', by a frequency correction that depends on the instantaneous orbital parameters. The various conservative corrections contained in \(\delta \Omega \) can be derived by enforcing that asymptotic observables \cite{tanaka} are consistent with post--Newtonian results in the weak field \cite{blanchet}. Once these corrections have been implemented the orbital frequency takes the form \cite{cons},
\begin{eqnarray}
\frac{\mathrm{d}\phi}{\mathrm{d}t} \equiv \Omega &=&\pm \frac{\sqrt{M}}{p^{3/2} \pm a \sqrt{M}} \Bigg(1 + \delta \Omega \Bigg),  \nonumber\\ 
 & =& \pm \frac{\sqrt{M}}{p^{3/2} \pm a \sqrt{M}} \Bigg\{1 +  \eta \left( d_0 + d_1 \left(\frac{M}{p}\right)+ (d_{1.5} + q\, l_{1.5})\left(\frac{M}{p}\right)^{3/2} + d_2\left(\frac{M}{p}\right)^{2}\right)\Bigg\}, \label{omCC}
\end{eqnarray}
\noindent where the various coefficients are given by 
\begin{eqnarray}
d_0 = \frac{1}{8}, \qquad d_1 = \frac{1975}{896}, \qquad d_{1.5}= -\frac{27}{10} \pi, \qquad l_{1.5}= -\frac{191}{160}, \qquad d_2 = \frac{1 152 343}{451 584}. 
\label{1.5}
\end{eqnarray}
\noindent  These are the ingredients that we use to build the inspiral part of our waveform model. In this paper, we will not explore the relative importance of the various terms that enter these expressions but leave that exercise for future work.
 
Once the inspiral trajectory has been computed, the inspiral waveform can be obtained from an expansion of the form
\begin{equation}
h(t) = -(h_{+} - i h_{\times}) = \sum_{\ell=2}^{\infty} \sum_{m=-\ell}^{l} h^{\ell m} {}_{-\!2}Y_{\ell m}(\theta,\Phi),
\label{inspwav}
\end{equation}
\noindent where the spin--weight \(-2\)  spherical harmonics  \(_{-2}Y_{\ell m}(\theta,\Phi)\) are given in terms of the Wigner \(d\) functions by
\begin{eqnarray}
_{-s}Y_{\ell m}(\theta,\Phi)&=& (-1)^s \sqrt{\frac{2\ell+1}{4 \pi}}d^{\ell}_{ms}(\theta)e^{im\Phi}, \\
\mbox{with }\qquad d^{\ell}_{ms}(\theta)&=&\sqrt{(\ell+m)!(\ell-m)!(\ell+s)!(\ell-s)!} \sum_{k=k_i}^{k_f} \frac{(-1)^k(\sin\frac{\theta}{2})^{2k+s-m}(\cos\frac{\theta}{2})^{2\ell+m-s-2k}}{k!(\ell+m-k)!(\ell-s-k)!(s-m+k)!},
\label{spinw}
\end{eqnarray}
\noindent where \(k_i={\rm max}(0,m-s)\) and \(k_f={\rm min}(\ell+m,\ell-s)\). Additionally, the complex conjugates of the spin--weighted spherical harmonics satisfy
\begin{equation}
_{s}Y^{\ell m * }(\theta,\Phi) = (-1)^{s+m}\, _{-s}Y^{\ell\, -m  }(\theta,\Phi).
\label{conj}
\end{equation}
\noindent We include only the modes \((\ell,m)=(2, \pm 2)\) in our model, which means that the components of the waveform at leading order are
\begin{eqnarray}
h_{+}(t)&=& \frac{4 \mu \left[\Omega(t) \, p(t)\right]^2}{D}\left(\frac{1+\cos^2 \theta}{2}\right)\cos\left[2(\phi(t) + \Phi)\right],\label{inspp}\\
h_{\times}(t)&=& \frac{4 \mu \left[\Omega(t) \, p(t)\right]^2}{D} \cos\theta \sin\left[2(\phi(t) + \Phi)\right],
\label{inspc}
\end{eqnarray}
\noindent where  \(\Omega(t) = d\phi/dt\) is the orbital frequency, \(p(t)\) is the radius of the orbit, and \(D\) is the distance to the source.

This model provides a description of the inspiral only, but it ceases to be valid when the CO approaches the innermost stable orbit and adiabaticity begins to break down. We require a different model for the evolution of the CO through the inner--most stable circular orbit (ISCO) all the way down to the light ring, where we match on to ringdown radiation.

\section{Transition and plunge phases for an initially spinning IMBH}
\label{s2b}
The approach we adopt to describe the merger waveform is based on the paper of Ori \& Thorne \cite{amos}, which describes the transition from inspiral to plunge for circular-equatorial inspirals  into a massive spinning black hole in the extreme-mass-ratio limit. To be consistent with their notation, we re-express the Boyer--Lindquist coordinates \((t,p,\theta,\phi)\) in dimensionless form using the transformations \(\tilde r = p/M\) and \(\tilde t = t/M\).

As described in section~\ref{s2.2}, during the inspiral phase, the CO moves on a circular geodesic orbit with dimensionless angular velocity
\begin{equation}
\tilde\Omega \equiv 
M\Omega = {\rmd\phi\over \rmd\tilde t} = {1\over{\tilde r^{3/2}+q}}.
\label{lessmega}
\end{equation}
\noindent As the CO inspirals onto the spinning IMBH, it radiates energy which is carried away by GWs. The radiation flux can be written as
\begin{equation}
\dot E_{\rm GW}  = - \dot E
= {32\over 5} \eta^2\tilde\Omega^{10/3}\dot {\cal E},
\label{dotE}
\end{equation}
\noindent with \(\dot {\cal E}\) being a general relativistic correction to the Newtonian, quadrupole-moment formula. 

The adiabatic inspiral phase evolution continues until the CO approaches the ISCO. As we discussed in the previous section, the adiabatic prescription breaks down somewhat before ISCO, as the orbit starts to evolve more quickly and the instantaneous-geodesic approximation is no longer valid. Hence, we need to find some point near the ISCO, \(\tilde r_{\rm trans} \gtrsim \tilde r_{\rm ISCO}\), that joins smoothly the adiabatic inspiral of Section~\ref{s2.2} onto the transition phase. This choice must ensure a continuous and smooth matching of the inspiral and transition waveforms. The transition radius, \(\tilde r_{\rm trans} \), is the point at which  \(\dot \tilde r_{\rm inspiral} \) becomes ``too fast''. ``Too fast'' obviously has a different meaning according to the binary system under consideration as it depends on the mass--ratio \(\eta\). However, the transition solution described in the following provides a unique solution that should match smoothly onto the inspiral. In practice, we chose a fixed matching radius for each binary system in order to give a smooth matching onto the transition waveform, but have subsequently verified that changing the value of \(\dot \tilde r_{\rm inspiral}\) at which the matching is done  does not significantly alter the resulting waveform, i.e., our choice of \(\tilde r_{\rm trans}\) is robust.

At the transition point, the circular geodesic has dimensionless angular velocity, energy and angular momentum given by
\begin{equation}
\tilde\Omega_{\rm trans} \equiv M\Omega
= {1\over {\tilde r_{\rm trans}}^{3/2} +q}\;,
\label{Omegams}
\end{equation}
\begin{equation}
\tilde E_{\rm trans} \equiv {E_{\rm trans}\over\mu} = {E_{\rm trans}\over\eta M}
= {1-2/\tilde r_{\rm trans} + q/{\tilde r_{\rm trans}}^{3/2}\over
\sqrt{1-3/\tilde r_{\rm trans} + 2q/{\tilde r_{\rm trans}}^{3/2} } }\;,
\label{Ems}
\end{equation}
\begin{equation}
\tilde L_{\rm trans} \equiv {L_{\rm trans}\over\mu M} = \tilde r_{\rm trans}^{1/2}\frac{1  - 2q/\tilde r_{\rm trans}^{3/2} + q^2/\tilde r_{\rm trans}^2}{\sqrt{1 - 3/\tilde r_{\rm trans} + 2q/\tilde r_{\rm trans}^{3/2}}}.
\label{lmstrans}
\end{equation}
The values of these quantities for the binary systems under consideration are given in Table~\ref{transitionII}. We have included the values of these quantities evaluated at ISCO for reference.

\begin{table}[thb]
\begin{tabular}{|c|c|c|c|c|c|c|}
\hline\multicolumn{1}{|c|}{}&\multicolumn{3}{c|}{$q=0.9$}&\multicolumn{3}{c|}{$q=0.3$}\\\cline{2-7}
\multicolumn{1}{|c|}{Binary systems}&$\tilde{L_{\mathrm{z}}}$&$\tilde{E}$&$\tilde r_{\rm trans}$&$\tilde{L_{\mathrm{z}}}$&$\tilde{E}$&$\tilde r_{\rm trans}$\\\hline
[10+100] $M_{\odot}$      &2.106&0.8477&2.450&3.157&0.9309&5.124\\\cline{1-7}
[1.4+100] $M_{\odot}$     &2.104&0.8472&2.442&3.155&0.9308&5.120\\\cline{1-7}
ISCO                     &2.100&0.8442&2.321&3.154&0.9306&4.979\\\cline{1-7}
[10+500] $M_{\odot}$      &2.104&0.8472&2.428&3.155&0.9308&5.075\\\cline{1-7}
[1.4+500] $M_{\odot}$     &2.104&0.8472&2.427&3.155&0.9308&5.075\\\hline
\end{tabular}
\caption{Dimensionless values for the energy $\tilde{E}$ and angular momentum $\tilde{L_{\mathrm{z}}}$ as defined in Eq.\eqref{1.1} at the point of transition \(\tilde r_{\rm trans}\). The values for the energy and angular momentum at ISCO have been included for reference. }
\label{transitionII}
\end{table}

\noindent As the CO enters the transition regime, the geodesic motion ceases to be adiabatic but radiation reaction continues to drive the orbital evolution. In this regime the CO still moves on a nearly circular orbit with radius close to  \(\tilde r_{\rm trans} \). Additionally, since the radiation reaction is proportional to the mass ratio and is therefore weak, the angular velocity and proper time can be approximated by \cite{amos}
\begin{eqnarray}
{\rmd\phi\over \rmd\tilde t} &\equiv& \tilde\Omega \simeq {\tilde\Omega_{\rm trans}}\;,
\label{dphidt} \\
{\rmd\tilde\tau\over \rmd\tilde t} &\simeq& 
\left({\rmd\tilde\tau\over \rmd\tilde t}\right)_{\rm trans} =
{ \sqrt{1-3/\tilde r_{\rm trans} + 2q/{\tilde r_{\rm trans}}^{3/2}}\over 1+q/{{\tilde r_{\rm trans}}}^{3/2}}\;.
\label{dtaudt}
\end{eqnarray}
\noindent In the vicinity of \(\tilde r_{\rm trans} \), the CO's energy and angular momentum can be written 
\begin{equation}
\tilde E = \tilde E_{\rm trans} + \tilde \Omega_{\rm trans} \xi\;,\quad 
\tilde L = \tilde L_{\rm trans} + \xi\;,
\label{defxi}
\end{equation}
\noindent where
\begin{eqnarray}
{\rmd\xi\over \rmd\tilde \tau} &=& -\kappa\eta\;, \quad \rm{and} 
\label{dxidtau}\\
\kappa &=& {32\over5}{\tilde\Omega_{\rm trans}}^{7/3}
{1+q/{\tilde r_{\rm trans}}^{3/2}\over\sqrt{1-3/\tilde r_{\rm trans} + 2q/{\tilde r_{\rm trans}}^{3/2}}}\dot {\cal E}_{\rm trans}\;.
\label{kappa}
\end{eqnarray}
\noindent To make further progress, we recast the effective potential describing the radial motion for geodesics
\begin{eqnarray}
V(\tilde r, \tilde E, \tilde L) &=& \tilde E^2 - 
{1\over \tilde r^4}\left([\tilde
E(\tilde r^2+q^2)-\tilde L q]^2
-(\tilde r^2-2\tilde r +q^2)[\tilde r^2+(\tilde L-\tilde E q)^2]
\right)\;,
\label{VrEL}
\end{eqnarray}
\noindent as a function of \(\tilde r\) and \(\xi \equiv \tilde L - \tilde L_{\rm trans}\). We use the variable \(R \equiv \tilde r-\tilde r_{\rm trans}\) to parametrise the CO's location during the transition regime. Both \(R\) and \(\xi\) are small and therefore we can expand the effective potential in terms of these two variables \begin{equation}
V(R,\xi) = {2\alpha\over3}R^3-2\beta R\xi +\hbox{constant}\,,
\label{VRxi}
\end{equation}
\noindent where \(\alpha\) and \(\beta\) are constants to be computed below. This new expression for the effective potential can be plugged into the equation describing the radial evolution in the transition regime, namely,
\begin{equation}
{\rmd^2 \tilde r\over \rmd\tilde\tau^2} = -{1\over2}{\partial V(\tilde r,
\xi)\over\partial \tilde r} + \eta\tilde F_{\rm self}\;.
\label{eom}
\end{equation}
\noindent The radial self--force  \(\eta\tilde F_{\rm self}\) is approximately non--dissipative and hence can be ignored. Absorbing this term into \(-{1\over2}\partial V/\partial\tilde r\) effectively amounts to changes in \(\tilde r_{\rm trans}\), \(\tilde E_{\rm trans}\), \(\tilde L_{\rm trans}\) and \(\alpha\) by fractional amounts proportional to \(\eta\).  Likewise,  these various quantities change by order \({\rm O}(\eta)\) as a result of the CO's perturbation of the black hole's spacetime geometry \cite{inner}. We shall ignore these small corrections in the following analysis.

Once we insert the effective potential \eqref{VRxi} into the equation of motion \eqref{eom} and set \(\tilde\tau \equiv 0\) at the moment when \(\xi=0\), we obtain \cite{amos}
\begin{equation}
{\rmd^2 R\over \rmd\tilde\tau^2} = -\alpha R^2-\eta\beta\kappa\tilde\tau\;,
\label{eom2}
\end{equation}
\noindent where the constants \(\alpha\, , \beta\) are given by
\begin{equation}
\alpha = {1\over4}
\left({\partial^3 V(\tilde r, \tilde E, \tilde L)\over\partial 
\tilde r^3}\right)_{\rm trans}\;,
\label{alpha}
\end{equation}
\begin{equation}
\beta = - {1\over2}
\left( {\partial^2 V(\tilde r, \tilde E, \tilde L)\over 
\partial \tilde L \partial \tilde r} + \tilde\Omega 
{\partial^2 V(\tilde r, \tilde E, \tilde L)\over 
\partial \tilde E \partial \tilde r} \right)_{\rm trans}\;.
\label{beta}
\end{equation}
\noindent  Eventually the transition regime ends, radiation reaction is no longer important and pure plunge takes over. Thereafter, the CO plunges towards the black hole with nearly constant energy and angular momentum given by \cite{amos}
\begin{eqnarray}
\tilde L_{\rm fin}-\tilde L_{\rm trans}&=&-(\kappa \tau_0 T_{\rm plunge}) \eta^{4/5}\;, \nonumber\\
\tilde E_{\rm fin}-\tilde E_{\rm trans}&=&- \tilde \Omega_{\rm trans}(\kappa \tau_0 T_{\rm plunge}) \eta^{4/5}\;,
\label{final-values}
\end{eqnarray}
\noindent where,
\begin{equation}
T_{\rm plunge} = 3.412\;, \qquad \tau_o = (\alpha\beta\kappa)^{-1/5}\;.
\label{Tplunge}
\end{equation}
\noindent During the plunge phase, the evolution is given by the Kerr geodesic equations~\cite{mtw}, i.e.,
\begin{eqnarray}
{\rmd^2 \tilde r\over \rmd\tilde\tau^2}&=&\frac{6\, \tilde E_{\rm fin}\, \tilde L_{\rm fin}\,q  + \tilde L^2_{\rm fin}\,(\tilde r-3) +(q^2-\tilde r)\tilde r -\tilde E^2_{\rm fin}\, q^2(\tilde r+3)}{\tilde r^4},
\label{rplun}\\
{\rmd\phi \over \rmd\tilde t}&=&\frac{\tilde L_{\rm fin}\,(\tilde r -2) + 2\,\tilde E_{\rm fin}\, q }{\tilde E_{\rm fin}\,(\tilde r^3 +(2+\tilde r)\, q^2 ) - 2\, q\, \tilde L_{\rm fin}},
\label{phiplu}\\
{\rmd\tilde \tau \over \rmd\tilde t}&=&\frac{\tilde r \,(q^2+ \tilde r\, (\tilde r -2))}{\tilde E_{\rm fin}\,(\tilde r^3 +(2+\tilde r)\, q^2 ) - 2\, q\, \tilde L_{\rm fin}}.
\label{tau}
\end{eqnarray}
\noindent Notice that the plunge angular frequency \eqref{phiplu} is entirely determined by the values of the energy and angular momentum \eqref{final-values}. Hence, to match the transition regime onto the plunge regime, we only need to find the point \(\tilde r_{\rm plunge}\) at which the transition angular frequency \eqref{lessmega} and the plunge angular frequency \eqref{phiplu} smoothly match for these specific values of energy and angular momentum \eqref{final-values}.

We now have all the ingredients required to build our waveform model from inspiral to plunge. For the inspiral waveform the cross and plus polarizations are given by \eqref{inspp}, \eqref{inspc}, which provide a good approximation before \(\tilde r_{\rm trans}\). Thereafter the particle is no longer on a circular orbit. Hence, we shall use the following expressions
\begin{eqnarray}
h_{+}(t)&=& \frac{\mu}{2D} \Big[\Big\{1 - 2\cos2\theta\cos^2[ \phi(t)+ \Phi] - 3\cos[2(\phi(t)+\Phi)]\Big\}\dot{ r^2} \nonumber \\ &+&
     (3 + \cos2\theta)\Big\{2\cos[2(\phi(t)+\Phi)]\dot{\phi^2(t)} + \sin[2(\phi(t)+ \Phi)]\ddot{\phi(t)}\Big\}r^2  + 
   \Big\{4(3 + \cos2\theta)\sin[2(\phi(t)+\Phi)]\dot{\phi(t)}\dot{r} 
   \nonumber \\ &+&  (1 - 2\cos2\theta\cos^2[\phi(t)+ \Phi] - 3\cos[2(\phi(t)+\Phi)])\ddot{r}\Big\}r \Big],
\label{inspptrans}\\
h_{\times}(t)&=&  \frac{-2 \mu \cos\theta}{D}\Big[ \sin[2(\phi(t)+\Phi)]\dot{r^2} + 
     \Big\{\cos[2(\phi(t)+\Phi)]\ddot{\phi(t)} -2\sin[2(\phi(t)+\Phi)]\dot{\phi^2(t)}  \Big\} r^2\nonumber \\  &+& 
     \Big\{ 4\cos[2(\phi(t)+\Phi)]\dot{\phi(t)}\dot{r} + \sin[2(\phi(t)+\Phi)]\ddot{r} \Big\}r \Big],
\label{inspctrans}
\end{eqnarray}
\noindent where \(\mu\) is the mass of the CO and \(D\) is the distance to the source. These are flat-spacetime emission formulae applied to a geodesic in curved space and so is in keeping with the ``numerical kludge'' approach to waveform generation. This approximation is vindicated by the similarity of the results to the EOB waveforms which we will demonstrate later.

This ``transition''  waveform model provides a consistent modelling for the gravitational radiation emitted from the early stages of inspiral evolution all the way down to the horizon. However, we shall attach the final part of the waveform, i.e., the RD part, at the effective light ring. The following section describes the method for attaching the plunge phase onto a RD waveform. The method is generic, i.e, it is applicable both for ``transition'' model, and for the EOB scheme which we will describe in Section~\ref{s2.3}.

\section{Ringdown waveform}
\label{s2c}
The ringdown radiation originates from the distorted Kerr black hole that is the end product of the merger, and consists of a superposition of quasinormal modes (QNMs), labelled by indices \((\ell,m,n)\), where \((\ell,m)\) specifies the mode and \(n\) the tone. Each mode has a complex frequency \(\hat \omega\), whose real part is the oscillation frequency and whose imaginary part is the inverse of the damping time \cite{rdspin},
\begin{equation}
\hat \omega=\omega_{\ell mn}- i/\tau_{\ell mn}.
\label{rdome}
\end{equation}
\noindent These two observables are uniquely determined by the mass and angular momentum of the newly formed Kerr black hole. Recent numerical simulations have shown that the total mass-energy radiated during the merger of two equal--mass maximally spinning BHs ranges from \(0.6\%-5\%\) of the total rest mass energy. The energy released in IMRI ringdown radiation will be much lower, as it is suppressed by the mass ratio $\eta$.

It would be reasonable to ignore the change in mass of the IMBH and approximate the final mass and spin by the initial values for the central black hole. But, for completeness, we will use a one--parameter fit, derived within the framework of the  EOB formalism, as an approximation for the final mass \(M_f\) of the black hole even though this does not account for the IMBH spin \cite{fai},

\begin{equation}
 M_f/M = 1 + (\sqrt{8/9}-1) \eta -0.498 (\pm 0.027)\, \eta^2. 
\label{finm}
\end{equation}

\noindent This fit is consistent with NR simulations to about \(\sim 2 \%\) accuracy for mass ratios \(\eta \gtrsim 0.16\) \cite{fspin}. However, the extrapolation to smaller values of \(\eta\) is consistent with NR simulations \cite{unequal} and test--mass limit predictions \cite{fspin}. Note that the coefficient of the linear term has been fixed to the test--mass limit value.

We will also use the fit by Rezzolla, et. al. \cite{spif}, to compute the final spin of the black hole
\begin{equation}
a_f/M_f = q_f = q + s_4\, q^2\, \eta + s_5\, q\, \eta^2 + t_0\, q\, \eta + 2\, \sqrt{3}\, \eta + t_2\, \eta^2 + t_3\, \eta^3,
\label{spn}
\end{equation}

\noindent in which the coefficients, obtained via a least--squares fit to available data, are
\begin{eqnarray}
s_4 = -0.129 &\pm& 0.012, \qquad  s_5 =0.384 \pm 0.261, \nonumber\\
t_0 = -2.686 &\pm& 0.065, \qquad  t_2 =-3.454 \pm 0.132, \qquad t_3 =2.353 \pm 0.548 .
\end{eqnarray} 

\noindent This fit was derived by using available data for spin parameters \(q \lesssim 0.8\) and mass ratios \(\eta \gtrsim 0.16\) along with exact results which hold in the extreme--mass--ratio limit, i.e., \(\eta \rightarrow 0\). For \(q=1\) the fit \eqref{spn} is a non--monotonic function with maximum \(q_f=1.029\) for \(\eta\simeq 0.093\), but this fit should be valid for the systems we consider here, for which \(\eta \lesssim 0.08\)  and \(q\leq0.9\). At present we do not have a good idea as to plausible values of \(q\) for the systems that will be detected by ET. If the dominant process by which IMBHs acquire spin is through the capture of COs, their angular momenta will undergo a damped random walk \cite{cole}, \cite{bland}.  This process was studied in detail by Mandel~\cite{mandel}, who computed the probability distribution for the spin of IMBHs that gain mass following a series of minor mergers. This work suggested that IMBHs with masses in the \(\sim 10^2\)--\(10^4 M_{\odot}\) range would have spin parameter \(q \sim 0.3\). 

Given the mass and spin of the final Kerr BH, we can uniquely determine the complex ringdown frequencies \eqref{rdome}.  We do this by  building an interpolation function based on the data provided in Table 2 of Berti, et. al. \cite{rdeq}, and evaluating this function for the final IMBH spin. Following Berti, et. al. \cite{rdeq}, and Buonanno, et. al. \cite{rdspin}, we construct a RD waveform that includes the fundamental mode \((\ell=2,m=2,n=0) \) and two overtones \((n=1,2)\). In order to be consistent, we also include the ``twin'' modes with frequency \(\omega'_{\ell mn}=-\omega_{\ell -mn}\) and a different damping \(\tau_{\ell mn}'=\tau_{\ell-mn}\). We include the ``twin modes'' since a mode with a given \((\ell,m)\) will always consist of a superposition of two different exponentials.  It might be the case that one of the exponentials has a shorter damping time or is less excited in the given physical situation and hence become ``invisible'', but formally we cannot have an isolated ``\(\ell=m=2\)'' frequency with a positive real part. Furthermore, a single--mode expansion restricts attention to circularly polarized GWs. These considerations are not important for non--spinning BHs as the two mirror solutions are then degenerate in the modulus of the frequency and in the damping time. Adding black hole rotation acts in a similar way to an external magnetic field on the energy levels of an atom, causing a Zeeman splitting effect of the QNM frequencies. The ringdown waveform is given by
\begin{eqnarray}
h(t) = -(h_+-i h_\times) &=& \frac{M_{f}}{D}\sum_{lmn} \biggl\{{\cal A}_{\ell mn} e^{-i(\omega_{\ell mn}t+\phi_{\ell mn})}e^{-t/\tau_{\ell mn}} S_{\ell m}(a\omega_{\ell mn})\nonumber\\&+&
{\cal A}'_{\ell mn} e^{i(\omega_{\ell mn}t+\phi'_{\ell mn})}e^{-t/\tau_{\ell mn}} S_{\ell m}^{*}(a\omega_{\ell mn}) \biggr\}.
\label{genwave}
\end{eqnarray}
\noindent  in which \(M_f\) and \(D\) are the mass of the Kerr BH formed after merger and the distance to the source, respectively.

The spherical harmonics of spin--weight \(-2\), $_{-2}S_{lm}$ obey the equation 
\begin{eqnarray}
& &\Bigl[{1 \over \sin\theta}{d \over d\theta}
 \Bigl\{\sin\theta {d \over d\theta} \Bigr\}
-a^2\omega^2\sin^2\theta
 -{(m-2\cos\theta)^2 \over \sin^2\theta}
+4a\omega\cos\theta-2+2ma\omega+\lambda\Bigr] 
{}_{-2}S_{\ell m}(a\omega)=0. \label{eq:spheroid2}
\end{eqnarray}

\noindent Expanding \(_{-2}S_{\ell m}(a\omega)\) and the eigenvalue \(\lambda\)~\cite{mine}
\begin{eqnarray} 
{}_{-2}S_{\ell m}(a\omega)
&=&{}_{-2}Y_{\ell m}+a\omega S_{\ell m}^{(1)}
+(a\omega)^2 S_{\ell m}^{(2)}
+O((a\omega)^3), \nonumber\\
\lambda&=&\lambda_0+a\omega\lambda_1
+a^2\omega^2 \lambda_2+O((a\omega)^3), \label{eq:slmexp2}
\end{eqnarray}
\noindent where \({}_{-2}Y_{\ell m}\) are the spherical harmonics of spin weight \(s=-2\). The normalizations of \({}_{-2}Y_{\ell m}\) and \({}_{-2}S_{\ell m}(a\omega)\) are fixed by
\begin{equation}
\int_0^{\pi} |_{-2}Y_{\ell m}|^2 \sin\theta d\theta
=\int_0^{\pi} |_{-2}S_{\ell m}(a\omega)|^2 \sin\theta d\theta=1.
\end{equation}
\noindent After plugging \eqref{eq:slmexp2} into \eqref{eq:spheroid2} and collecting terms, we obtain the zero and first order corrections to the eigenvalue \(\lambda\), i.e.,
\begin{displaymath}
\lambda_0=(\ell-1)(\ell+2), \qquad \lambda_1=-2m{\ell(\ell+1)+4\over\ell(\ell+1)}\,.
\end{displaymath}
\noindent The first order correction to  \(_{-2}S_{\ell m}(a\omega)\) is given by
\begin{equation}
S_{\ell m}^{(1)}=\sum_{\ell'} c_{\ell m}^{\ell'}~{}_{-2}Y_{\ell'm}\,,
\label{eq:slm11}
\end{equation}
\noindent where the non-zero coefficients \(c_{\ell m}^{\ell'}\) are 
\begin{eqnarray}
c_{\ell m}^{\ell+1}&=&{2 \over (\ell+1)^2}\Bigl\lbrack 
{(\ell+3)(\ell-1)(\ell+m+1)(\ell-m+1) \over (2\ell+1)(2\ell+3)}
 \Bigr\rbrack^{1/2},\cr
c_{\ell m}^{\ell-1}&=&-{2 \over \ell^2}\Bigl\lbrack 
{(\ell+2)(\ell-2)(\ell+m)(\ell-m) \over (2\ell+1)(2\ell-1)} 
\Bigr\rbrack^{1/2}.
\end{eqnarray} 
\noindent  This first order approximation will suffice for our analysis since \(a\omega \ll 1 \) in Eq. \eqref{eq:spheroid2} for the systems we shall consider. Using these relations we can split \eqref{genwave} into plus and cross polarizations, which can be matched onto the corresponding components of the plunge waveforms. This matching requires the determination of 24 constants, 12 for each polarization, but the equations can be rewritten as two 6D systems of equations. We do this matching at the time that the particle reaches the light-ring.

The approach to determine the various amplitudes and RD phases of Eq. \eqref{genwave} is the following: use the plunge waveform to compute ten points before and after the LR to build an interpolation function. This function can be used to match onto the various QNMs by imposing continuity of the waveform and all the necessary higher order time derivatives. We first match the plunge waveform onto the leading RD tone \(n=0\) at the point where the orbital frequency (\eqref{phiplu}, \eqref{eob:1}) peaks, \(t_{\rm peak}\). This fixes 2 constants per polarization. We then use these values as a seed to compute the amplitudes and phases of the first overtone at a time \( t_{\rm peak} + dt \). Finally, we use the values of the amplitudes and phases of the leading tone and first overtone to determine the four remaining constants at a time  \( t_{\rm peak} + 2dt \).

In terms of the relative SNRs contributed by the different tones,  we have found that the leading tone along with its twin mode are the major contributors. These two modes contribute more than \(90 \%\) of the SNR for the various binary systems under consideration. The remaining two overtones and their twin modes do not contribute substantially to SNR but we have included them for completeness and to achieve the best possible modelling of the RD waveform.

\section{Plunge and merger waveform from the effective one-body approach}
\label{s2.3}

In order to cross--check the predictions of our IMRI ``transition model'' in the non--spinning limit, we have used an independent approach to model the plunge and merger phases, which is based on the Effective-One-Body (EOB) formalism. Note, however, that the the ringdown waveform presented in Section~\ref{s2c} is generic and can be applied to this new prescription. Therefore, within this framework, a complete waveform model consists of the inspiral waveform described in Section~\ref{s2.2}, the plunge and merger waveform to be discussed below, and the ringdown waveform described in Section~\ref{s2c}. 

Although we use the EOB model only in the non-spinning limit, an extension of the EOB scheme does exist which includes leading--order spin--orbit and spin--spin dynamical effects of a binary system for an ``effective test particle'' moving in a Kerr--type metric \cite{eobs}, and next--to--leading--order spin--orbit couplings \cite{ntlo}. However, it has been recently found \cite{gsve} that it is not straightforward to include higher--order non--spinning PN couplings, such as the 4PN and 5PN adjustable parameters that were recently calibrated to numerical relativity simulations for non-spinning systems~\cite{dna, eobnon}, using these Hamiltonians~\cite{eobs,ntlo}. Additionally, the EOB Hamiltonian in \cite{ntlo} does not reduce to the Hamiltonian of a spinning test particle in Kerr spacetime. This issue was recently resolved in \cite{cha}, in which a canonical Hamiltonian was derived for a spinning test particle in a generic curved spacetime at linear order in the particle spin. The construction of an improved EOB Hamiltonian based on the results of \cite{cha} has recently been obtained \cite{hams}.  The Hamiltonian derived in \cite{ntlo} has recently been used in an exploratory study to calibrate the EOB parameters using numerical relativity simulations of spinning, non--precessing, equal mass BHs.  This is the same approach that was previously used with great success for non-spinning black hole systems, e.g., to derive fits for the final mass and spin of a BH after merger that are consistent with NR to about \(\sim 2 \%\) accuracy~\cite{fspin}. We used the non-spinning EOB model only in this work because that model is more mature. However, the EOB model has recently been used to model circular--equatorial extreme--mass--ratio inspirals (EMRIs) around spinning supermassive black holes, using fits of various post--Newtonian parameters to Teukolsky--based waveforms~\cite{nic}. Comparisons between an IMRI model based on this spinning EOB framework and the waveforms constructed using the transition model should be pursued in the future.

The standard analytic method to study the two-body dynamics of comparable--mass BHs is the PN framework, which is an expansion in the characteristic orbital velocity \(v/c\). Accurate equations of motion at 2.5PN order were derived in the 1980's \cite{pn25}. At present, corrections at 3.5PN--level in the equations of motion are available \cite{35pnw,kon,nis} and the diffeomorphism--invariant dimensional regularization method proposed by Blanchet, et al., \cite{reg} opened up the way to also study the GWs emitted by inspiralling non--spinning compact binaries up to 3.5PN order of accuracy. The PN formalism is reliable as long as the PN expansion parameter \(\delta =M/d \ll 1/6 \) , where \(M\) is the total mass of the binary system and \(d\) is the separation between the two BHs, i.e., only during the early inspiral stage. Once \(\delta \gtrsim 1/12 \), we require an alternative description of the motion and radiation to accurately model the final stages of inspiral plus merger and ringdown~\cite{rev}. Numerical Relativity (NR) is the best candidate to model these various evolution phases for comparable--mass BHs. However, the need to compute thousands of waveform templates to carry out matched filtering searches makes NR an impractical tool due to the high computational cost of producing individual waveforms. 

In order to circumvent this problem, Buonanno \& Damour \cite{eobz} introduced a framework to study the entire waveform using conservative dynamics at 2PN order. This scheme provides an accurate model from the early stages of inspiral down to the LSO. Inside the LSO, the scheme includes a non--perturbative plunge down to the LR and, thereafter, the model can be matched onto ringdown radiation. The basic claim is that it is possible to use analytical tools to obtain a sufficiently accurate waveform from inspiral to RD. Instead of using PN expansions in the original form, resummation methods (a non--polynomial function of the mass ratio \(\eta=m_1m_2/M^2\), which incorporates some of the expected non--perturbative features of the exact results) can be used to improve the convergence properties of the PN expansions. In the test--mass limit (\(\eta \rightarrow 0\)) the two basic ingredients of a GW signal, the two--body energy and the GW energy flux, are first resummed. The resulting EOB model is constructed to exactly recover geodesic motion in the test--mass limit. The argument was that the resummed quantities would also provide a good description for the comparable--mass case, since this is in effect a smooth deformation of the test-mass limit.  The intrinsic flexibility of the model allowed a natural extension to higher order \cite{tres} once the 3PN calculation was completed. This belief has been vindicated by recent advances in numerical relativity, as it has proven possible to construct EOB waveforms that match the results of the numerical simulations very well \cite{rev}. For instance, the EOB model is the only analytic approach that has successfully predicted the spin of the final BH after a merger to better than \(\sim 2 \%\) accuracy \cite{fspin}.

In building the model, one of the challenges was to encode the conservative part of the relative orbital dynamics into the dynamics of an effective particle, i.e., to map the \textit{ real} conservative two--body dynamics at the highest PN order available, onto an \textit{effective} one--body problem, in which a test particle of mass \(\mu = m_1 m_2 /M\) moves in some effective background metric \(g_{\mu \nu} ^{\rm eff} \). In the particular case of non--spinning binaries and ignoring radiation--reaction effects, the best effective metric was found to be a deformation of the Schwarzschild metric, with \(\eta\) playing the role of a deformation parameter.

We shall now introduce the mathematical machinery required to construct the plunge waveform model. We first show how to obtain the cross and plus polarizations for the plunge phase evolution and then explain how we match the inspiral waveform model of Section~\ref{s2.2} onto this plunge waveform. We shall use phase space variables \((r,\phi, p_r,p_\phi)\) to write the effective EOB Hamiltonian for non--spinning binaries at 3PN order as follows
\begin{widetext}
\begin{equation} 
\label{eq:genexp}
H_{\rm eff}(\mathbf{r},\mathbf p) = \mu\, 
\widehat{H}_{\rm eff}({\mathbf r},{\mathbf p}) 
= \mu\,\sqrt{A (r) \left[ 1 + 
{\mathbf p}^{2} +
\left( \frac{A(r)}{D(r)} - 1 \right) ({\mathbf n} \cdot {\mathbf p})^2
+ \frac{1}{r^{2}} \left( z_1 ({\mathbf p}^{2})^2 + z_2 \, {\mathbf p}^{2}
({\mathbf n} \cdot {\mathbf p})^2 + z_3 ({\mathbf n} \cdot {\mathbf p})^4 \right) \right]} \,,
\end{equation}
\end{widetext}
\noindent where \(\mathbf{n} = \mathbf{r}/r\) and  \(r = |{\mathbf{r}}|\). It is convenient to replace the radial momentum \(p_r\) by the momentum conjugate to the tortoise radial coordinate \(r_{*} = \int dr \left(B(r)/A(r)\right)^{1/2}\), where \(A(r), \, B(r)\) are metric functions which will be defined below.  We do this because \(p_{r_{*}}\) tends to a finite value after merger, whereas \(p_r\) diverges at the event horizon. Such a coordinate transformation allows a more controlled treatment of the late part of the EOB dynamics \cite{rev}. Under this coordinate transformation the EOB Hamiltonian takes the form
\begin{equation}
\label{eqn11}
\hat H_{\rm eff} = \sqrt{p_{r_*}^2 + A(r) \left( 1 + \frac{p_{\phi}^2}{r^2} + z_3 \, \frac{p_{r_*}^4}{r^2} \right)} \, ,
\end{equation}
\noindent and the mapping between the real and effective EOB Hamiltonian is given by
\begin{equation}
\label{eqn10}
\hat H_{\rm EOB} (r,p_{r_*} , \phi) = \frac{H_{\rm EOB}^{\rm real}}{\mu} 
= \dfrac{1}{\eta}\sqrt{1 + 2 \eta \, (\hat H_{\rm eff} - 1)} \, ,
\end{equation}
\noindent where \(\hat{H}_{\rm real} = H_{\rm real}/\mu\). This relation holds true at 2PN and 3PN order. The arbitrary coefficients  \(z_1,z_2\) and \(z_3\) in Eq.~\eqref{eq:genexp} are subject to the constraint
\begin{equation}
\label{cef}
8z_1 + 4z_2 +3z_3 = 6(4-3\eta)\,\eta\,.  
\end{equation}

\noindent It is possible to determine these coefficients by means of a fit to numerical results for comparable mass systems. We should bear in mind that in so doing, we must recover exact results in the test--mass limit. To achieve this, we have to ensure that \(z_1,z_2\) and \(z_3\) must go to zero as \(\eta \rightarrow 0\). On the other hand, Damour, et al., \cite{zed}, found that the terms proportional to \(z_2, z_3\) in Eq. \eqref{eq:genexp} are very small for quasicircular orbits. They also noticed the convenience of setting \(z_1=0\) because in any other case \(z_1\) could be suitably chosen so as to cancel the 3PN contribution in the metric coefficient \(A(r)\). Hence, we shall follow the general philosophy adopted by Damour et. al. \cite{zed}, and Buonanno et. al., \cite{cof} and set \({z}_1={z}_2=0\), \(z_3 = 2(4 - 3\eta)\eta\). In order to ensure the existence and \(\eta\)--continuity of a last stable orbit as well as the existence and \(\eta\)--continuity of an \(\eta\)--deformed analog of the light-ring, the metric coefficient \(A(r)\) must be Pad\'e resummed,
\begin{widetext}
\begin{equation}
\label{coeffPA}
A_{P_3^1}^{\rm 3PN}(r) = \frac{r^{2}\,[(a_4(\eta)+8\eta-16) + r\,(8-2\eta)]}{
r^{3}\,(8-2\eta)+r^{2}\,[a_4(\eta)+4\eta]+r\,[2a_4(\eta)+8\eta]
+4[\eta^2+a_4(\eta)]}\,,
\end{equation}
\end{widetext}
\noindent where 
\begin{equation} \label{a4}
a_4(\eta) = \left [\left (\frac{94}{3}-\frac{41}{32}\pi^2 \right )\,\eta \right ]\,.
\end{equation}
\noindent The light-ring in the test--mass limit is the solution to \(d/dr (A(r)/r^2 =0)\). For \(\eta \neq 0\), the \(\eta\)--deformed LR is obtained by solving  \(d/dr (A(r,\eta)/r^2) =0\). The existence of this ``deformed'' LR guarantees that in its vicinity the orbital frequency \(\Omega\) reaches a maximum. Additionally, the Pad\'e resummation of the metric coefficient \(D(r)\) ensures an \(\eta\)--continuity in the plunging phase. Its Pad\'e resummed form at 3PN is given by 

\begin{equation}
\label{D3PN}
D_{P_3^0}^{\rm 3PN}(r)=\frac{r^3}{r^3+6\eta r+2 \eta(26-3\eta)}\,.
\end{equation}
\noindent Using reduced quantities \(\hat{H}_{\rm real} = H_{\rm real}/\mu\), \(\hat{t} = t/M\), \(\hat{\Omega} = \Omega\,M\), the EOB equations of motion that describe the waveform evolution from the LSO to the LR are \cite{fai},

%
\begin{align}
%
\label{eob:1}
\dfrac{\rmd\phi}{\rmd\hat t}     &= \dfrac{A p_\phi}{\eta r^2\hat{H}\hat{H}_{\rm eff}} \equiv \hat\Omega\ , \\
\label{eob:2}
\dfrac{\rmd r}{\rmd\hat t}           &= \left(\dfrac{A}{B}\right)^{1/2}\dfrac{1}{\eta\hat{H}\hat{H}_{\rm eff}}\left(p_{r_*}+z_3\dfrac{2A}{r^2}p_{r_*}^3\right) \ , \\
\label{eob:3}
\dfrac{\rmd p_{\phi}}{\rmd\hat t} &= \hat{\cal F}_{\phi} \ , \\
\label{eob:4}
\dfrac{\rmd p_{r_*}}{\rmd\hat t}     &= -\left(\dfrac{A}{B}\right)^{1/2}\dfrac{1}{2\eta\hat{H}\hat{H}_{\rm eff}}
  \left\{A'+\dfrac{p_\phi^2}{r^2}\left(A'-\dfrac{2A}{r}\right)+z_3\left(\dfrac{A'}{r^2}-\dfrac{2A}{r^3}\right)p_{r_*}^4
  \right\} \ ,
\end{align}

\noindent where \(\hat{\cal F}_{\phi}\) is the \(\phi\) component of the radiation--reaction force.

We shall use this set of equations only to model the waveform evolution from the LSO to the LR and it is therefore reasonable to assume that the two--body dynamics is no longer driven by radiation--reaction but occurs along a geodesic with constant angular momentum \(p_{\phi}\) given by \cite{fai}
\begin{eqnarray}
 p_\phi^{2} &=& - \left[\frac{A'(u)}{(u^2\,A(u))'}\right]_{\rm LSO}, \label{eq:eobph} 
\end{eqnarray}
\noindent where \(u=1/r\) and the prime denotes \(d/d u\). The motion during plunge remains quasi--circular in the sense that \(p_r^2\) stays numerically small compared to \( p_\phi^2\). 

In the limit \(\eta \rightarrow 0\), we know that circular orbits in a Schwarzschild geometry satisfy the Kepler law \(\Omega^2 r^3 =1\). It is common to use \(v_{\Omega} \equiv \Omega^{1/3}\) or \(x_{\Omega} \equiv  \Omega^{2/3}\) to describe all PN corrections whether they are proportional to the square of the linear azimuthal velocity (\(v_{\phi} = \Omega r\)) or to the gravitational potential (\(U = 1/r\)). In order to generalise this to the case \(\eta \neq 0\), we introduce the functions 
\begin{eqnarray}
\psi(r, p_\phi) &=& \frac{2}{r^2}\left(\frac{d A(r)}{d r}\right)^{-1} \left[1+2\eta \left(\sqrt{A(r)\left(1+\frac{p_\phi^2}{r^2}\right)} -1\right) \right],
\label{psieq} \\
r_{\Omega} &=& r\,\psi(r,p_\phi), \label{rmod}
\end{eqnarray} 
\noindent  so that this modified radius \(r_{\Omega}\) is related to \(\Omega\) by the standard Kepler law \(\Omega^2 r_{\Omega}^3 =1\). This expression holds true during plunge, while the combination \(K=\Omega^2 r^3\) becomes of order \(0.5\) at the effective LR. We use this modified relation, \(\Omega^2 r_{\Omega}^3 =1\), to determine \(\Omega\) inside the LSO. Finally, we use Eqs. (4.38) and (4.39) of \cite{prcs} to determine the value of \(p_{r_{*}}\) at the LSO from which we can solve the set of Eqs. \eqref{eob:1}, \eqref{eob:2} and \eqref{eob:4} for the plunge phase. 

Once the orbital evolution has been determined, the EOB waveform is given in terms of spin--weight \(-2\)  spherical harmonics  \(_{-2}Y_{\ell m}(\theta,\Phi)\) through the relation 
\begin{equation}
h_{\ell m } \equiv -(h_+ - i h_\times)_{\ell m} = -\int d \Omega\, {}_{-\!2}Y^*_{\ell m}(\theta,\Phi)\,({h}_+ - i{h}_\times)\,.
\end{equation}
\noindent As for the inspiral, we shall include only the two modes \((\ell,m)=(2, \pm 2)\). At leading PN order the two modes \(h_{22}\) and \(h_{2 -2}\) are given by
\begin{eqnarray}
\label{22}
\frac{D}{M}h^{\rm EOB}_{22}(t)&=& -8\,\sqrt{\frac{\pi}{5}}\,\eta\,(r_{\Omega}(t)\,\Omega(t))^{2}\,\,F_{2 2}(t)e^{-2i\,\phi(t)}\,, \\
h^{\rm EOB}_{2 -2}(t) &=&  \,h_{22}^*(t)
\label{44}
\end{eqnarray}
\noindent where \(D\) is the distance to the source, \(M\) is the total mass of the binary system and \(\phi(t)\) is the binary orbital phase. The factor \(F_{22}\) is a resummed version of all the PN corrections and is given by
\begin{equation}
\label{f22}
F_{22}(t) = \hat{H}_{\rm eff}\, T_{22}(t)\, \rho_{22}^2(x(t))\, e^{{\rm i}\delta_{22}(t)}, 
\end{equation}

\noindent where \(x(t) = r^{-1}_{\Omega}\) and \(T_{\ell m}(t)\) is a resummed version of an infinite number of logarithmic terms that enter the transfer function between the near--zone and far--zone waveforms. These terms arise due to tail effects connected to the wave propagation in a Schwarzschild background of mass \(M_{\rm ADM}= H^{\rm real}_{\rm EOB}\) \cite{rev}. The factor \(\delta_{\ell m}\) is a supplementary phase which corrects the phase effects not included in the complex tail factor  \(T_{\ell m}\). 

Finally, in order to enhance the agreement between the EOB model and numerically computed waveforms near the end of inspiral and during the beginning of plunge, we have introduced the resummed quantity \(\rho_{\ell m}\), which enters the waveform only through its \(\ell\)--th power, \(\rho_{\ell m}^\ell\). Previous waveform models utilised a different PN improving factor \(F_{22}\) (see  Eq.~\eqref{f22}), namely, \(F_{22}(t) = \hat{H}_{\rm eff}\, T_{22}(t)\, f_{22}(x(t))\, e^{{\rm i}\delta_{22}(t)}\), where \( f_{22}\) is a PN--expanded amplitude factor.  The Taylor--expanded \(f_{\ell m}\)'s  produce results that are incompatible with numerical data close to the LSO. This problem arises because the \(f_{\ell m}\)'s have coefficients that grow linearly with \(\ell\), and these terms are problematic for the accuracy of PN--expansions, as shown in~\cite{resu}. Replacing \(f_{\ell m}\) by its \(\ell\)--th root \(\rho_{\ell m}= [f_{\ell m}]^{1/\ell}\) seems to be a cure for these accuracy problems and improves agreement with Numerical Relativity in the strong--field/fast--motion regime \cite{rev}. The explicit forms of the various quantities introduced above are \cite{nova,resu}, 
\begin{eqnarray}
T_{2 2} &=& \frac{\Gamma(3 -2 i\hat{\hat{k}})}{\Gamma(3)}e^{\pi\hat{\hat{k}}} e^{2{ i}\hat{\hat{k}}\log(2 k r_0)},\\
\delta_{22} &=& \dfrac{7}{3}H_{\rm real}\Omega + \dfrac{428}{105}\pi\left(H_{\rm  real}\Omega\right)^2 -24\eta x^{5/2},\\
\rho_{22}(x;\nu)&=& 1 +\left(\frac{55 \nu }{84}-\frac{43}{42}\right) x 
+\left(\frac{19583 \nu^2}{42336}-\frac{33025 \nu
}{21168}-\frac{20555}{10584}\right) x^2 \nonumber\\
&+&\left(\frac{10620745 \nu ^3}{39118464}-\frac{6292061 \nu ^2}{3259872}+\frac{41 \pi
   ^2 \nu }{192}-\frac{48993925 \nu }{9779616}-\frac{428}{105}
  \text{eulerlog}_{2}(x)+\frac{1556919113}{122245200}\right) x^3 \nonumber\\
&+&\left(\frac{9202}{2205}\text{eulerlog}_2(x)-\frac{387216563023}{160190110080}\right) x^4
+\left(\frac{439877}{55566}\text{eulerlog}_{2}(x)-\frac{16094530514677}{533967033600}\right)x^5+{\cal O}(x^6), \nonumber\\
\end{eqnarray}
\noindent in which
\begin{eqnarray}
\hat{\hat{k}}&\equiv &G H_{\rm real} m\Omega,\qquad k=m\Omega, \nonumber\\
r_0 &= &2M,\qquad \qquad \quad x(t) =\frac{1}{r_{\Omega}},\nonumber\\
\text{eulerlog}_{2}(x) &=& \gamma_{E} + 2 \log 2 + \frac{1}{2}\log x,\quad \rm{with} \quad \gamma_{E}=0.577215, \nonumber\\
\end{eqnarray}
\noindent with \(M\) being the total mass of the binary as before.

In order to facilitate the matching between the inspiral and plunge waveforms we rewrite the factor \(F_{22}(t) = G(t)\,e^{ i \epsilon(t)}\), where both \(G(t)\) and \(\epsilon(t)\) are real quantities. The waveform can then be written as
\begin{widetext}
\begin{eqnarray}
h(t) &\equiv& -(h_+ - i h_\times)\, \nonumber \\
&=&{}_{-2}Y_{22}(\theta,\Phi)\,h_{22}(t)  + {}_{-2}Y_{2-2}(\theta,\Phi)\,h_{2-2}(t),
\label{h}
\end{eqnarray}
\end{widetext}
\noindent and Eq.~\eqref{h} allows us to write the cross and plus waveform components for plunge and merger as
\begin{eqnarray}
h_{+}(t)&=& \frac{4 \mu G(t)\left[\Omega(t) \, r_{\Omega}(t)\right]^2}{D}\left(\frac{1+\cos^2 \theta}{2}\right)\cos\left[2(\phi(t) + \Phi - \frac{1}{2}\epsilon(t))\right],\label{eobp}\\
h_{\times}(t)&=& \frac{4 \mu G(t)\left[\Omega(t) \, r_{\Omega}(t)\right]^2}{D} \cos\theta \sin\left[2(\phi(t) + \Phi - \frac{1}{2}\epsilon(t))\right].
\label{eobc}
\end{eqnarray}
\noindent Now that we have found explicit expressions for the plus and cross polarizations of the plunge waveform, we must match these on to the inspiral waveform, which is achieved as follows: we already know that the scheme used to evolve the radial coordinate in the inspiral phase, Eq. \eqref{1.3},  begins to break down as the CO nears the ISCO. Hence, we first need study the behaviour of Eq. \eqref{1.3} near the ISCO to find out where the adiabatic approximation starts to break down. We have found that for the binary systems under consideration, the transition to plunge occurs at a point \(r_{\rm trans} \approxeq r_{\rm isco}\). In practice, the matching point \(r_{\rm trans}\) was chosen to ensure that: i) the transition from inspiral to plunge of the radial, Eqs. \eqref{1.3}, \eqref{eob:2},  and azimuthal, Eqs. \eqref{omCC}, \eqref{eob:1}, coordinates is smooth; ii) the transition from inspiral, Eqs.~\eqref{inspp}, \eqref{eobp}, to plunge, Eqs.~\eqref{inspc}, \eqref{eobc}, of the waveform is smooth; and iii) the choice of \(r_{\rm trans}\) is robust, i.e., conditions i) and ii) are met in a vicinity around the precise value of \(r_{\rm trans}\). Furthermore, we have verified that the phasing and the amplitude of the resulting waveform are insensitive to the exact choice of $r_{\rm trans}$. Table~\ref{transitionI} indicates the energy and angular momentum, as defined in Eq. \eqref{1.1}, at the transition point \(r_{\rm trans}\) compared to the corresponding values for a test-particle at ISCO.

\begin{table}[thb]
\begin{tabular}{|c|c|c|c|}
\hline\multicolumn{1}{|c|}{Binary systems}&\multicolumn{1}{c|}{$\tilde{L_{\mathrm{z}}}$}&\multicolumn{1}{c|}{$\tilde{E}$}&\multicolumn{1}{c|}{$\tilde r_{\rm trans} $}\\\hline
[10+100] $M_{\odot}$      &3.46544&0.9429&6.103\\\cline{1-4}
[1.4+100] $M_{\odot}$     &3.46497&0.9429&6.101\\\cline{1-4}
ISCO                     &3.46410&0.9428&6.000\\\cline{1-4}
[10+500] $M_{\odot}$      &3.46499&0.9429&6.084\\\cline{1-4}
[1.4+500] $M_{\odot}$     &3.46500&0.9429&6.083\\\hline
\end{tabular}
\caption{Dimensionless values for the energy $\tilde{E}$ and angular momentum $\tilde{L_{\mathrm{z}}}$ as defined in Eq.~\eqref{1.1} at the point of transition \(\tilde r_{\rm trans}\). The values for the energy and angular momentum at ISCO have been included for reference. }
\label{transitionI}
\end{table}

By construction, the plunge waveform is a good description of the waveform all the way to the event horizon. However, we shall attach a set of quasinormal RD modes (QNMs) at the effective light-ring as in the transition model case. Following Buonanno, et. al. \cite{fai}, we attach the RD modes at the time when the orbital frequency \eqref{eob:1} peaks. The frequency of these ringdown modes depends on the mass and spin of the newly formed Kerr black hole after merger. We use the following EOB--based fit for the final mass and spin \cite{fai}
\begin{eqnarray}
 M_f/M &=& 1 + (\sqrt{8/9}-1) \eta -0.498 (\pm 0.027)\, \eta^2, 
\label{finm1}\\
a_f/M_f = q_f &=& \sqrt{12} \eta -2.900 (\pm 0.065)\, \eta^2.
\label{fiteob}
\end{eqnarray}
\noindent We note that this one--parameter fit for the final spin of the post--merger Kerr BH, Eq.~\eqref{fiteob}, differs from the zero-spin limit of the spin-dependent expression used for the ``transition model'', Eq.~\eqref{spn}. The two expressions render similar results, but we will use Eq.~\eqref{fiteob} to estimate the final spin of the post--merger Kerr BH in our EOB waveform model. The reason for this is two-fold, firstly it ensures that we use a consistent EOB framework for this second model and secondly, it is the conservative approach. We will use the EOB model both as a consistency check of the transition model and to assess the level of uncertainty in our results. The latter is best accomplished by using this alternative fit to the final spin to make the waveform as different as possible within results available in the literature. We have verified that the waveform models are not particularly sensitive to this choice, and the level of consistency we quote later encodes this uncertainty. The matching onto QNMs is obtained as before by imposing continuity of the EOB waveform, Eqs.~\eqref{eobp} and \eqref{eobc}, and all the higher order time derivatives that are necessary to determine the amplitudes and phases of the leading RD tone, two overtones and their respective  twin modes. The strategy to carry out this matching was described in Section~\ref{s2c}.

\section{Dynamics of the waveform models }
\label{s2.5}
We shall now combine the ingredients of the preceding sections to discuss the orbital evolution and waveforms for our twelve test systems, which have mass ratios \(\eta=0.0192\, ( 10M_{\odot}+500 M_{\odot})\), \(\eta = 0.0028\, (1.4M_{\odot}+500 M_\odot )\), \(\eta = 0.0826 \, (10M_{\odot}+100 M_{\odot})\) and \(\eta=0.0136 \, (1.4M_{\odot}+100 M_{\odot})\), and the three different spin parameters, namely, \(q=0,0.3,0.9\). For the zero spin case, we will compare the EOB model to the transition model as a consistency check. The final ingredient we need to include in the model is the response function and the noise model for the ET detector. 

\subsection{Implementation of the response function}
The ET response may be written as
\begin{equation}
h_{\alpha}(t)= \frac{1}{ D} \Big[F_\alpha ^ +(t)h^ +(t) + F_\alpha ^ \times(t) h^ \times(t)\Big],
\label{14}
\end{equation}
\noindent where $\alpha = I,II$ refers to the two independent right--angle Michelson-like detectors. Any number of coplanar and colocated detectors with uncorrelated noise have an equivalent GW response to that of two right--angle interferometers offset by \(45^{\circ}\) to one another, and it is these putative detectors that we label as $I$ and $II$. The functions \(h^ {+\, , \times} (t)\) are the two independent polarizations of the gravitational  waveform. The antenna pattern functions \(F_\alpha ^{+ \times}\) are given by
\begin{eqnarray}  
F^{+}_{I} &=&  \frac{1}{2}(1+\cos^2\theta)\cos(2\phi)\cos(2\psi)
-\cos\theta\sin(2\phi)\sin(2\psi), \nonumber\\
F^{\times}_{I} &=& \frac{1}{2}(1+\cos^2\theta)\cos(2\phi)\sin(2\psi)
+\cos\theta\sin(2\phi)\cos(2\psi),
\label{16} \\
F^{+}_{II} &=&  \frac{1}{2}(1+\cos^2\theta)\sin(2\phi)\cos(2\psi)
+\cos\theta\cos(2\phi)\sin(2\psi), \nonumber\\
F^{\times}_{II} &=& \frac{1}{2}(1+\cos^2\theta)\sin(2\phi)\sin(2\psi)
-\cos\theta\cos(2\phi)\cos(2\psi).
\label{17}
\end{eqnarray}  
\noindent The various angles in the previous expressions represent, (a) the source's sky location in a detector based coordinate system (\(\theta,\phi\)), and (b) the polarization angle of the wavefront (\(\psi\)). These can be re--written in a fixed, ecliptic--based coordinate system. If we denote the source co--latitude and azimuth angles in the ecliptic coordinate system by  (\(\theta_S,\phi_S\)), and  the direction of the IMBH's spin \(\hat a\) by  (\(\theta_K,\phi_K\)), then we can use the expressions (5), (7b) and (64a)--(64c) of Apostolatos et. al.,  \cite{apos} to determine \(\theta(t), \, \psi(t), \, \phi(t)\).

If \(\Phi(t)\) denotes the phase of the waveform at the centre of the Earth, there is also a shift in the waveform phase due to the difference in location  of the detector. This can be included as a phase shift \cite{cutler}
\begin{equation}
\Phi(t)\to \Phi(t)+ 2 \frac{\mathrm{d}\phi}{\mathrm{d}t} R \sin\theta_S \cos[2\pi(t/T)-\phi_S],
\label{20}
\end{equation}
\noindent where  \(R= {R_{\oplus}/c}= 0.02125\, \textrm{s}\), and \(\mathrm{d}\phi/\mathrm{d}t\) is the azimuthal velocity of the orbit. This term is different for different detectors in a network, and encodes the time delay information that allows source triangulation using the network.

\subsection{Noise model}
The signal--to--noise ratio for a given waveform is determined by an integral in Fourier space, weighted by the power spectral density (PSD) of the detector. For a monochromatic source, we can approximate this integral as a time domain integral by defining a noise--weighted waveform 
\begin{equation}\label{30}
\hat h_{\alpha}(t) \equiv   \frac{h_{\alpha}(t)}{\sqrt{S_h\bigl(f(t)\bigr)}}, \qquad f(t) = \frac{1}{\pi}\frac{\mathrm{d}\phi}{\mathrm{d}t}.
\end{equation}
\noindent Where the PSD, \(S_h\bigl(f\bigr)\), is evaluated at the instantaneous gravitational wave frequency, \(f(t)\). This is a good approximation during the inspiral, during which the frequency is given by Eq.~\eqref{omCC}, and for the ringdown, during which the PSD factor takes the form \(S_h \left(f(t)\right)=S_h\left(f_{\ell m n}\right)\), where \(f_{\ell m n}= \omega_{\ell m n}/2\pi\) \cite{rdeq}. Since the distorted Kerr BH will predominantly be emitting GWs at the leading mode \(\ell=m=2, n=0\), we set  \(S_h \left(f(t)\right)=S_h\left(f_{2 2 0}\right)\) during the RD phase. In the merger and plunge phase, this approximation is not appropriate, but this phase is short and so we use the same approximation, taking the instantaneous frequency from \eqref{eob:1} in the EOB case and \eqref{phiplu} in the transition model. Treated in this way, there is a small jump in the frequency used to evaluate the PSD when the orbit passes from plunge to ringdown, so we use interpolation to ensure a smooth transition. In the regime where this approximation is used, the PSD is quite flat (cf. Figure \ref{figura0}) so we do not expect significant errors from using this approach. Furthermore, we have checked that this treatment of the detector noise generates results that are consistent with results computed directly in the frequency domain.

\subsection{Sample waveforms}

\begin{figure*}[!htp]
\centerline{
\includegraphics[width=0.5\textwidth, angle=0,clip]{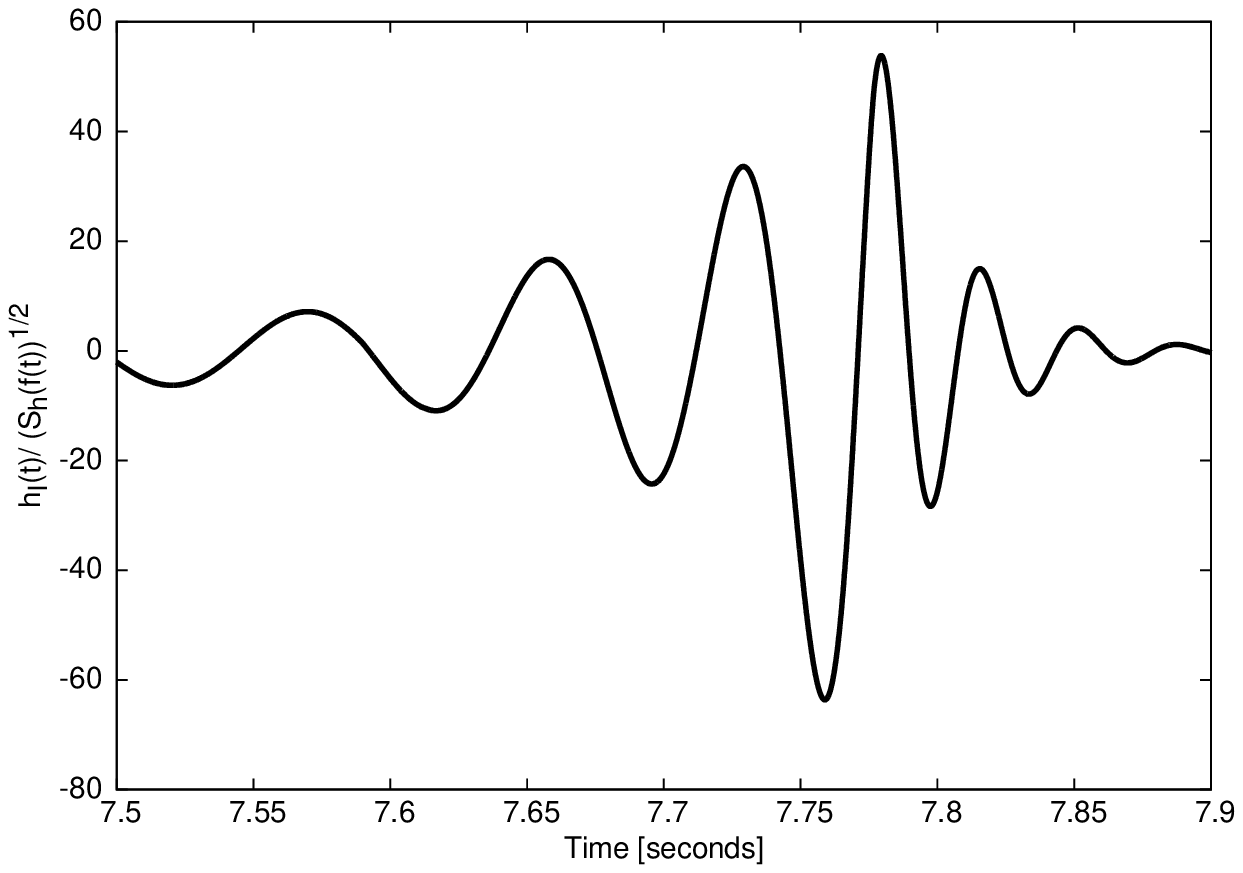}
\includegraphics[width=0.5\textwidth, angle=0,clip]{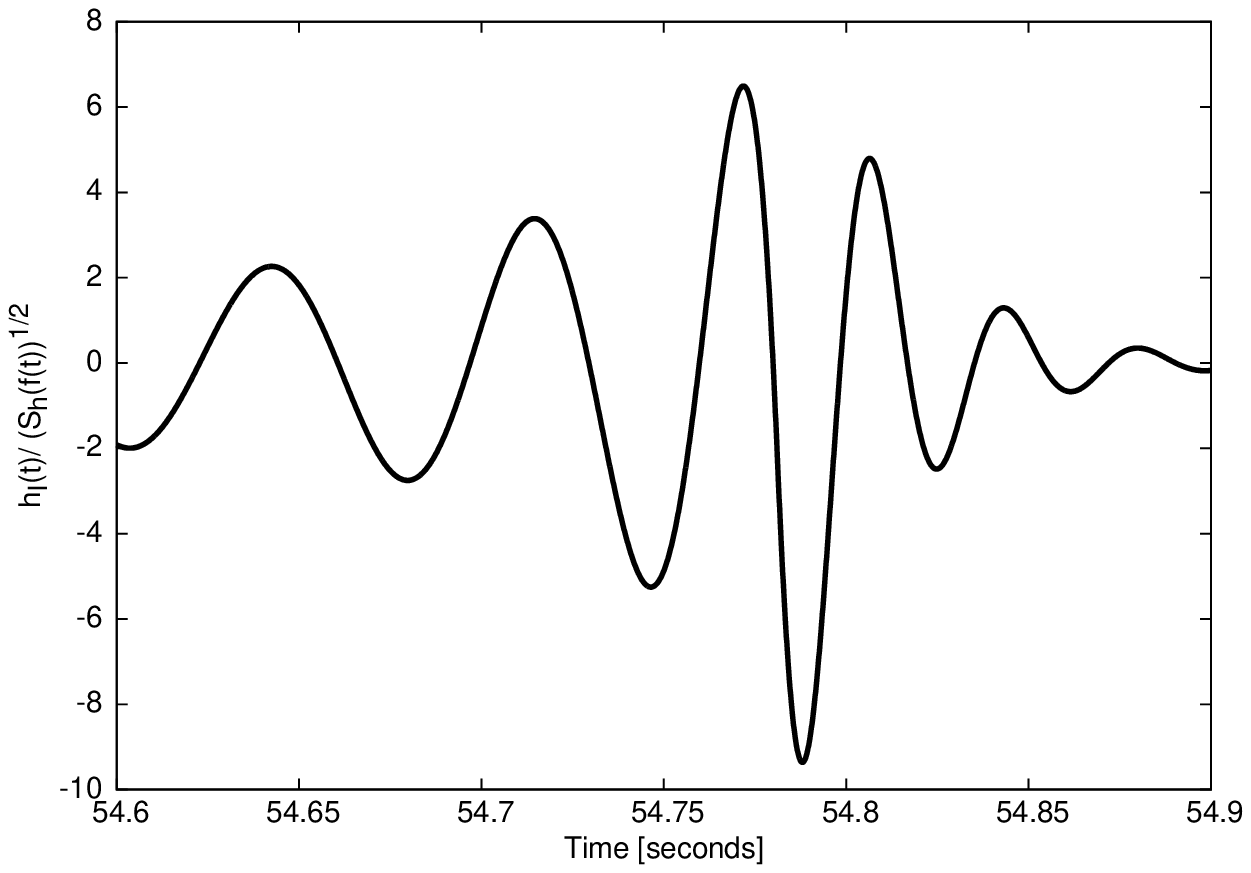}
}
\caption{Complete gravitational waveforms in the interferometer `I' for COs of masses \(10\,M_{\odot}\) (left panel) and \(1.4 M_{\odot}\) (right panel), orbiting around a \( 500 M_{\odot}\) BH with spin parameter \(q=0.3\). The gravitational waveform shows the last stage of inspiral, plus the transition, plunge and RD phases. The various extrinsic parameters were chosen randomly. }
\label{figura1}
\end{figure*}

We shall now put at work the machinery developed in Sections~\ref{s2.2}--\ref{s2.3} to generate a few sample noise--weighted waveforms. 
In  Figure \ref{figura1} we show the waveform for two sample systems. The waveforms in Figure \ref{figura1} have the expected form. We see a gradual chirping signal with increasing frequency and amplitude which peaks at the merger and is damped exponentially afterwards. The ringdown radiation is weaker for the lower mass inspirals, \(m=1.4 M_{\odot}\), as we would expect since the energy released scales as \(\eta^2\). 

\begin{figure*}[!htp]
\centerline{
\includegraphics[width=0.5\textwidth, angle=0,clip]{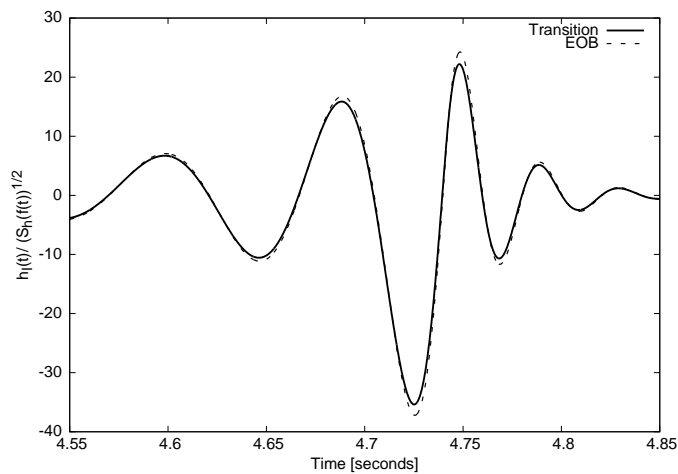}
}
\caption{ We show gravitational waveforms emitted during the final stage of inspiral, merger and ringdown for a \(10\,M_{\odot}\) CO orbiting around a \( 500 M_{\odot}\) BH, as computed using the EOB model and the transition waveform with \(q\) set to \(0\). As for Figure~\ref{figura1} the various extrinsic parameters were chosen randomly. }
\label{figura2}
\end{figure*}

In Figure~\ref{figura2} we compare the EOB and transition models for the waveform. Since both models were designed to yield accurate results in the test--mass limit, we expect a good agreement for small \(\eta\), but perhaps a larger deviation for large \(\eta\). In fact, the models agree well even for the larger value of \(\eta\). The phase is in very good agreement, which is important for matched filtering, and the amplitude agrees to better than \(10\%\). This provides confidence in our results, and the difference provides a guide to the overall level of uncertainty that we might expect in the SNRs which we quote in the next section. 

In the following section we present SNRs for these twelve different binary systems. The full parameter space of the waveform is ten dimensional and these parameters are defined in  Table \ref{tableparams}. For the SNR calculations, we fix the intrinsic parameters (the first four parameters of Table \ref{tableparams}) and the distance to the source and then run a Monte Carlo over the values of the remaining five extrinsic parameters.  

\begin{table}[thb]
\centerline{$\begin{array}{c|l}\hline\hline
 \ln m      & \text {mass of CO}   \\
 \ln M         & \text {mass of SMBH}\\
 q         & \text{magnitude of (specific) spin   angular momentum of SMBH} \\
t_0           & \text{time at which orbital frequency sweeps through a reference value} \\
\phi_0 &  \text{initial phase of CO orbit}      \\
\theta_S  &  \text{source sky colatitude in an ecliptic--based system }  \\
\phi_S   & \text{source sky azimuth in an ecliptic--based system}  \\
 \theta_K   & \text{direction  of SMBH spin (colatitude)}  \\
\phi_K       &  \text{direction of SMBH spin (azimuth)}  \\
\ln D          & \text{distance to  source}\\
\hline\hline
\end{array}$}
\caption{\protect\footnotesize
This table shows the physical meaning of the parameters used in our model.
The various angles ($\theta_S$,\,$\phi_S$) and ($\theta_K$,\,$\phi_K$) are defined in an ecliptic--based coordinate system.}
\label{tableparams}
\end{table}

\section{SNRs for spinning and non--spinning binaries}
\label{s2.6}

We start with a brief review of the framework of signal analysis. A GW detector can be thought of as a linear system whose input is a GW and whose output is a time series. This time series is a combination of both instrumental noise and a true GW signal. For ET, the output of each of the equivalent Michelson detectors can be represented as 
\begin{equation}
s_{\alpha}(t) = h_{\alpha}(t) + n_{\alpha}(t), \qquad \alpha = \textrm{I, \,II}.
\label{21}
\end{equation}
\noindent The detection problem is to distinguish  \( h_{\alpha}(t)\) from \( n_{\alpha}(t)\). We can define  
\begin{equation}
\hat s = \int_{-\infty}^\infty s(t) K(t) \, {\rm d}t,
\label{filter}
\end{equation}
\noindent where \(K(t)\) is a filter function. Given the fact that we have a model for the signal  \(h(t)\), we want to find the filter function that maximizes the SNR for such a GW signal.  This optimal filter function is usually known as Wiener filter.  There is a natural inner product on the vector space of signals, which for any two signals \(p_\alpha (t), q_\alpha(t)\), takes the form
\begin{equation}\label{inp}
\left( {\bf p} \,|\, {\bf q} \right)
\equiv 2\sum_{\alpha} \int_0^{\infty}\left[ \tilde p_\alpha^*(f)
\tilde q_\alpha(f) + \tilde p_\alpha(f) \tilde q_\alpha^*(f)\right]
/S_h(f)\,\mathrm{d}f,
\end{equation}
and in terms of this inner product, the maximal SNR obtained with the Weiner optimal filter is 
\begin{equation}
\frac{S}{N}[h(\theta^i)]= \frac{\left( {\bf s}\,|\, {\bf h} \right)}{\sqrt{\left( {\bf h}\,|\, {\bf h} \right)}},
\label{26}
\end{equation}
\noindent where the expected waveform \(h(t;\theta)\) depends on parameters \(\theta=\{\theta_1,...,\theta_N\}\).

\begin{figure*}[!htp]
\centerline{
\includegraphics[width=0.55\textwidth, angle=0, clip]{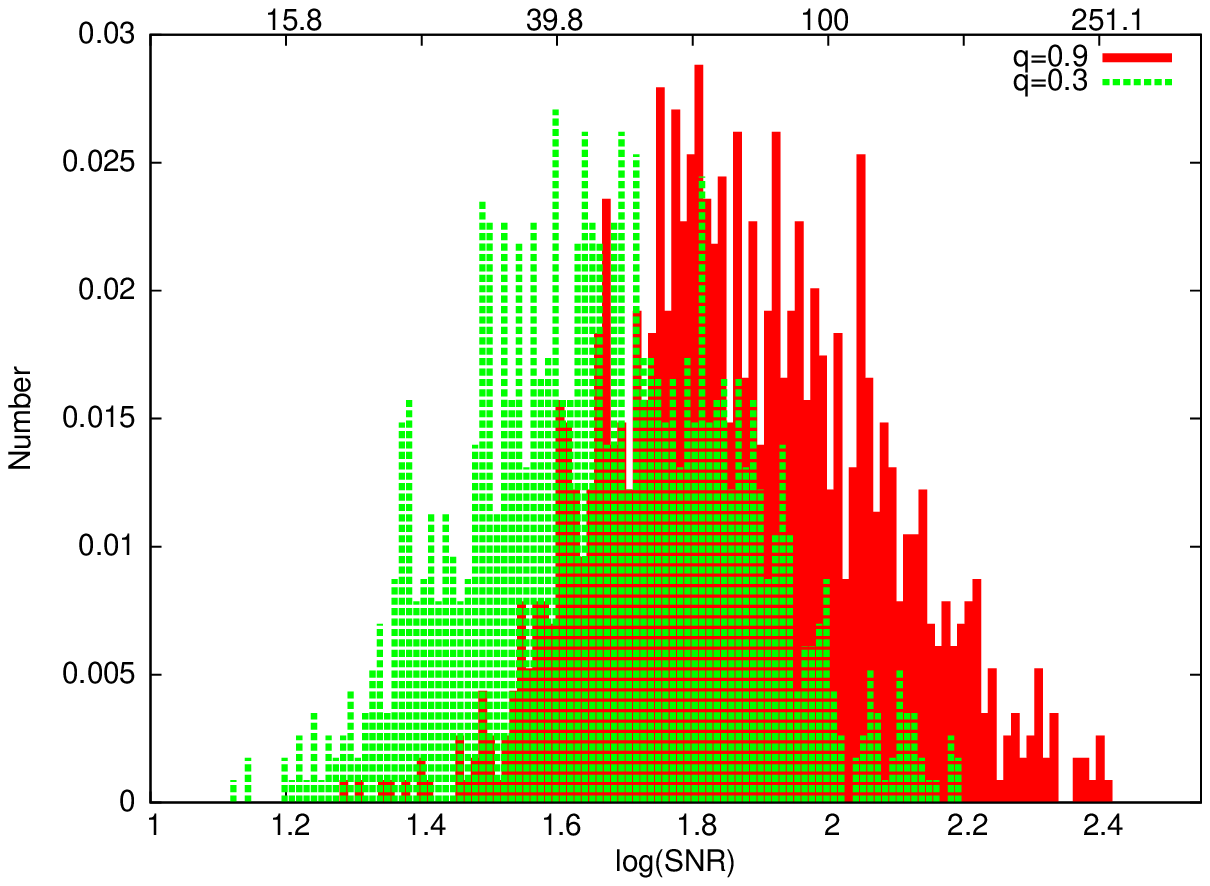}
\includegraphics[width=0.55\textwidth, angle=0, clip]{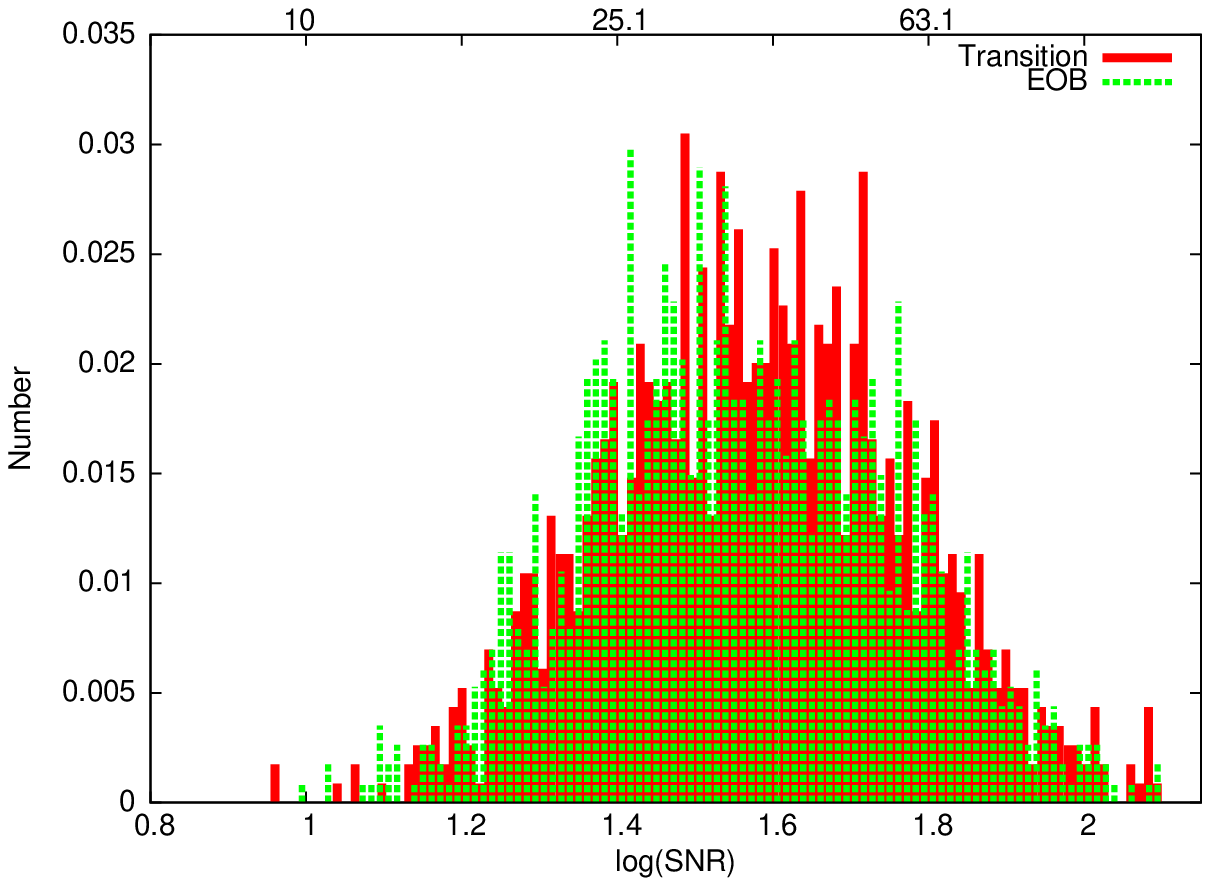}
}
\caption{ The left panel shows the distributions of SNRs in a Monte Carlo simulation over extrinsic parameters for a \(10\,M_{\odot}+100M_{\odot}\) binary with IMBH spin parameter \(q=0.9, \, 0.3\). The right panel shows the SNR distribution for the same binary system using the transition waveform model in the \(q=0\) limit along with its  EOB counterpart. The horizontal axis is the logarithm to base ten of the SNR. Note that the SNR distributions have been computed at a fixed distance \(D=6.63481\) Gpc, or, equivalently, at a fixed redshift \(z=1\). The top horizontal axis mirrors the values used in the bottom horizontal one, but in normal units.}
\label{figura3}
\end{figure*}

\begin{figure*}[!htp]
\centerline{
\includegraphics[width=0.55\textwidth, angle=0, clip]{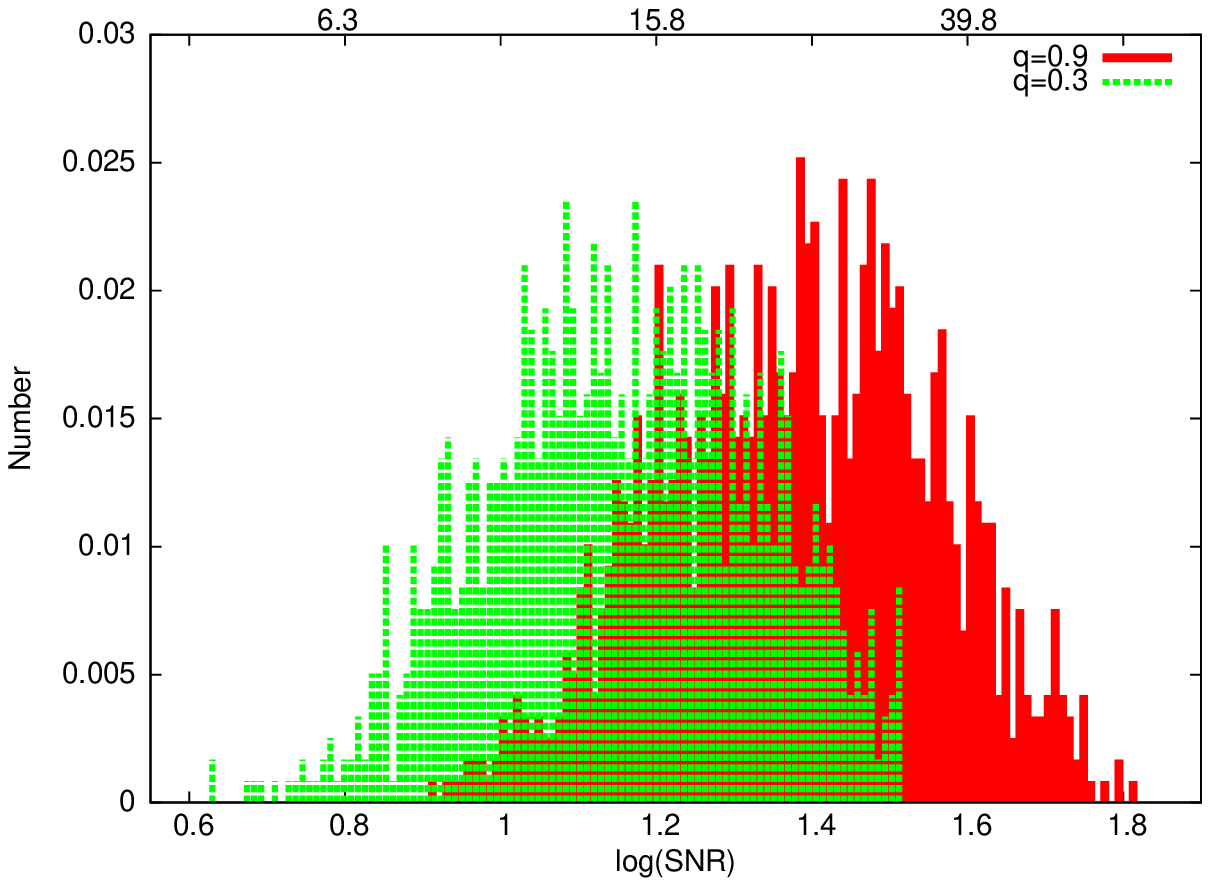}
\includegraphics[width=0.55\textwidth, angle=0, clip]{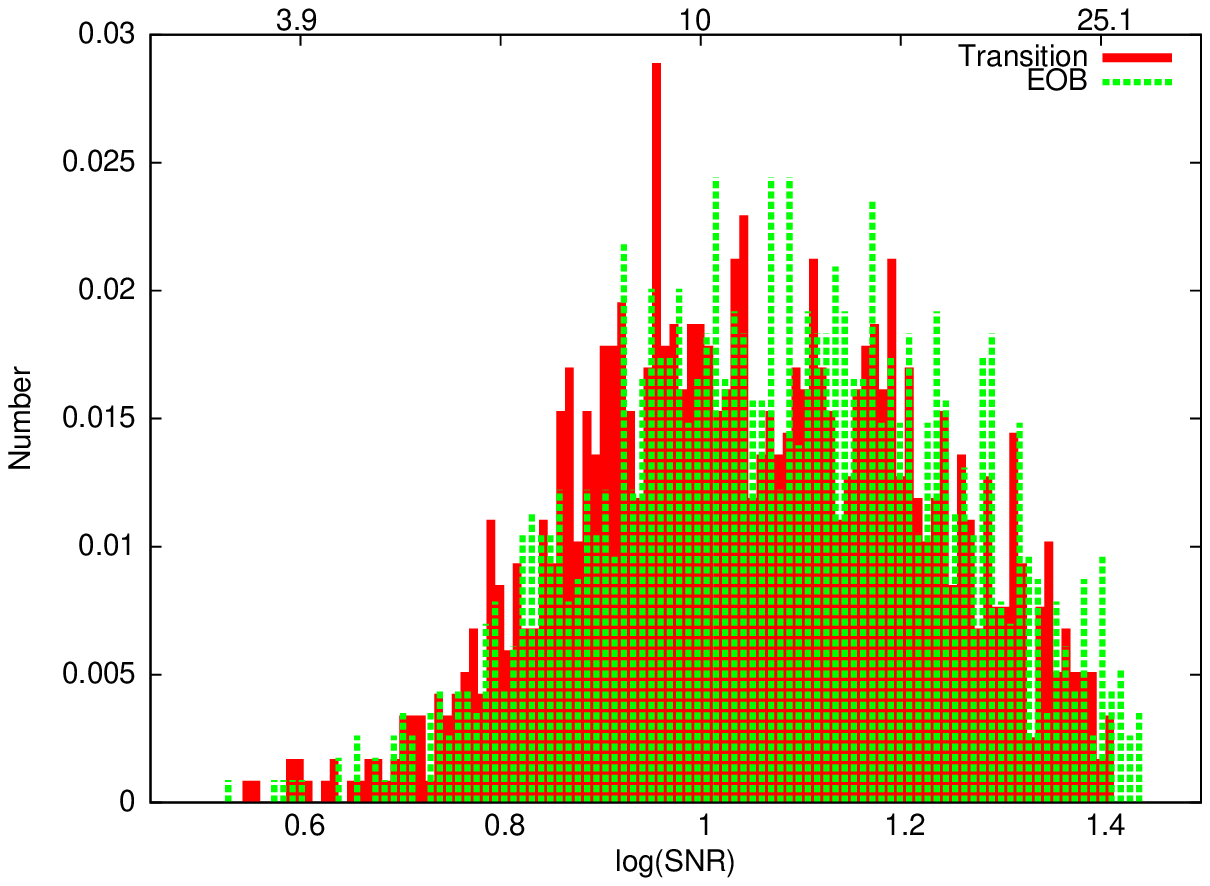}
}
\caption{ As Figure~\ref{figura3}, but now for binaries with masses \(1.4\,M_{\odot} + 100 M_{\odot}\). The left panel shows SNR distributions for IMBH spin $q=0.3$ and $q=0.9$, while the right panel shows SNR distributions for IMBH spin $q=0$ computed using both the EOB and transition models. 
}
\label{figura4}
\end{figure*}

\begin{figure*}[!htp]
\centerline{
\includegraphics[width=0.55\textwidth, angle=0, clip]{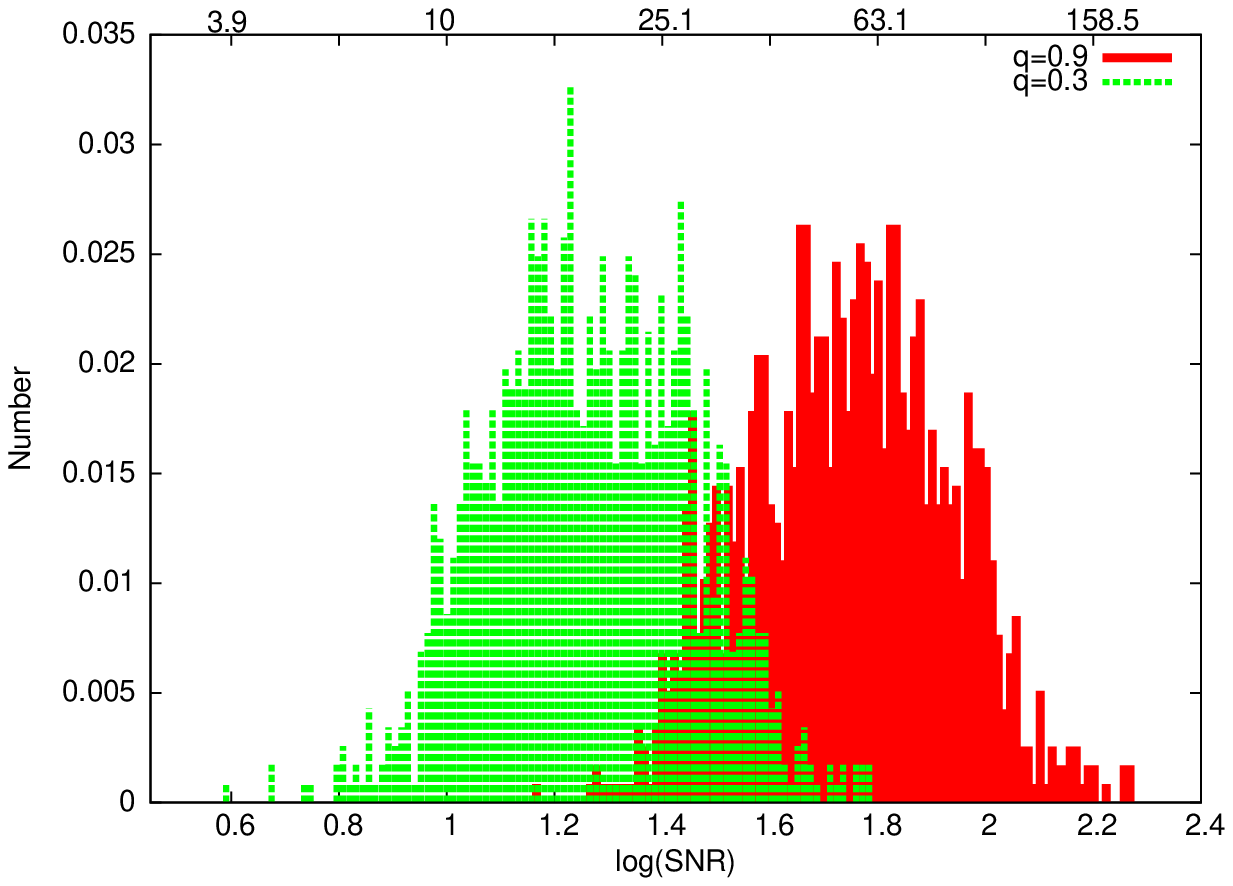}
\includegraphics[width=0.55\textwidth, angle=0, clip]{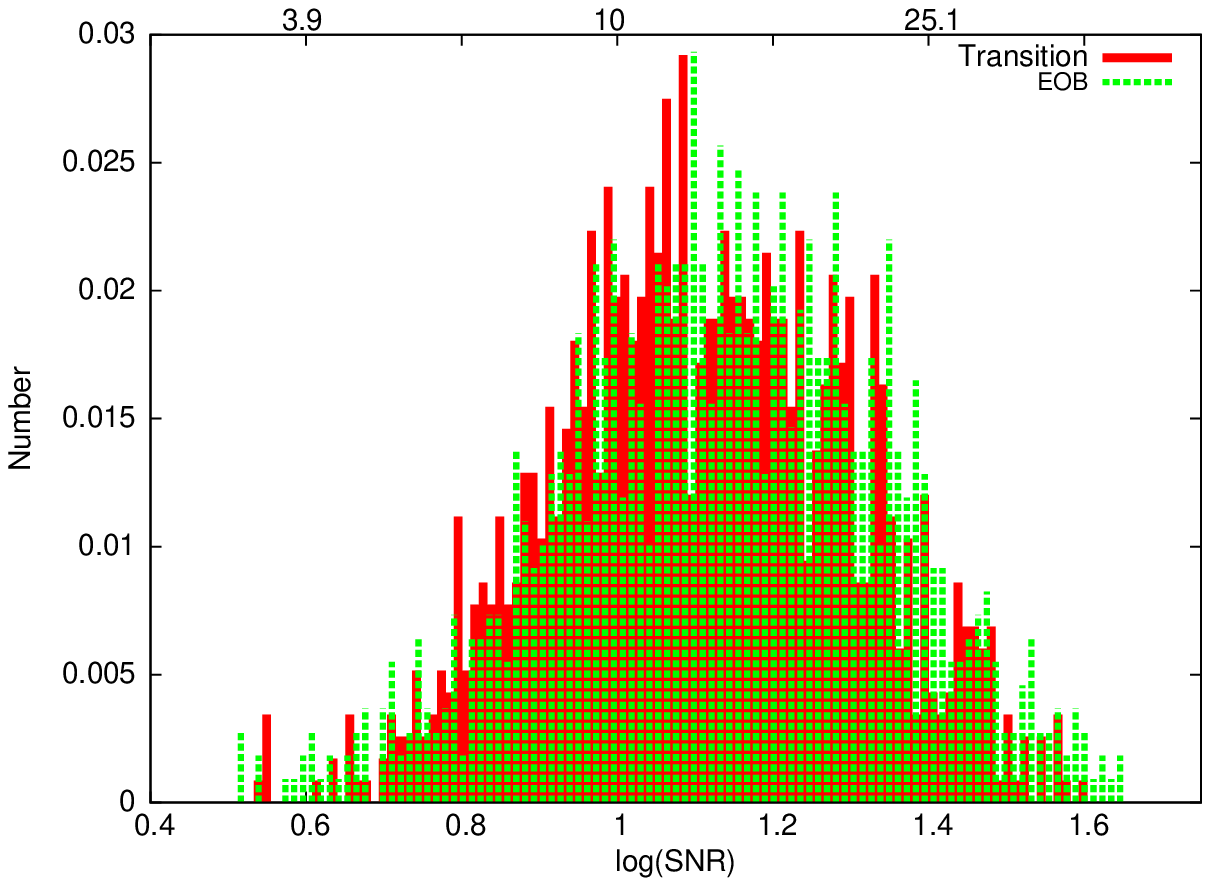}
}
\caption{ The left panel shows the SNR distributions for a \(10\,M_{\odot}\) CO orbiting around a \( 500 M_{\odot}\) BH with spin parameters \(q=0.9, \, 0.3\), respectively. The right panel shows the SNR distributions for the same binary system with $q=0$ using both waveform models. The panels show the distribution of the logarithm to base ten of the SNR. The SNR distributions have been computed at a fixed distance \(D=6.63481\) Gpc.  The top horizontal axis mirrors the values used in the bottom horizontal one, but in normal units. }
\label{figura5}
\end{figure*}

\begin{figure*}[!htp]
\centerline{
\includegraphics[width=0.55\textwidth, angle=0, clip]{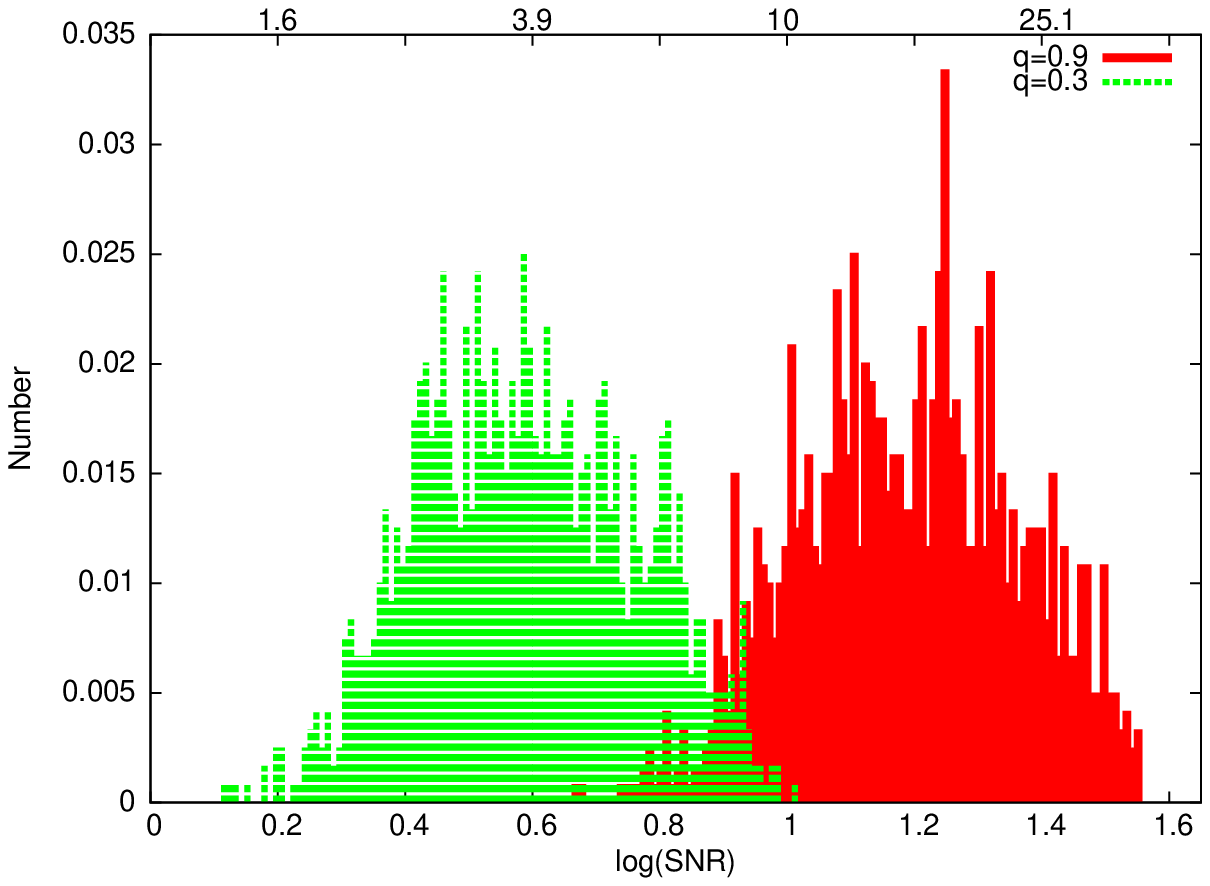}
\includegraphics[width=0.55\textwidth, angle=0, clip]{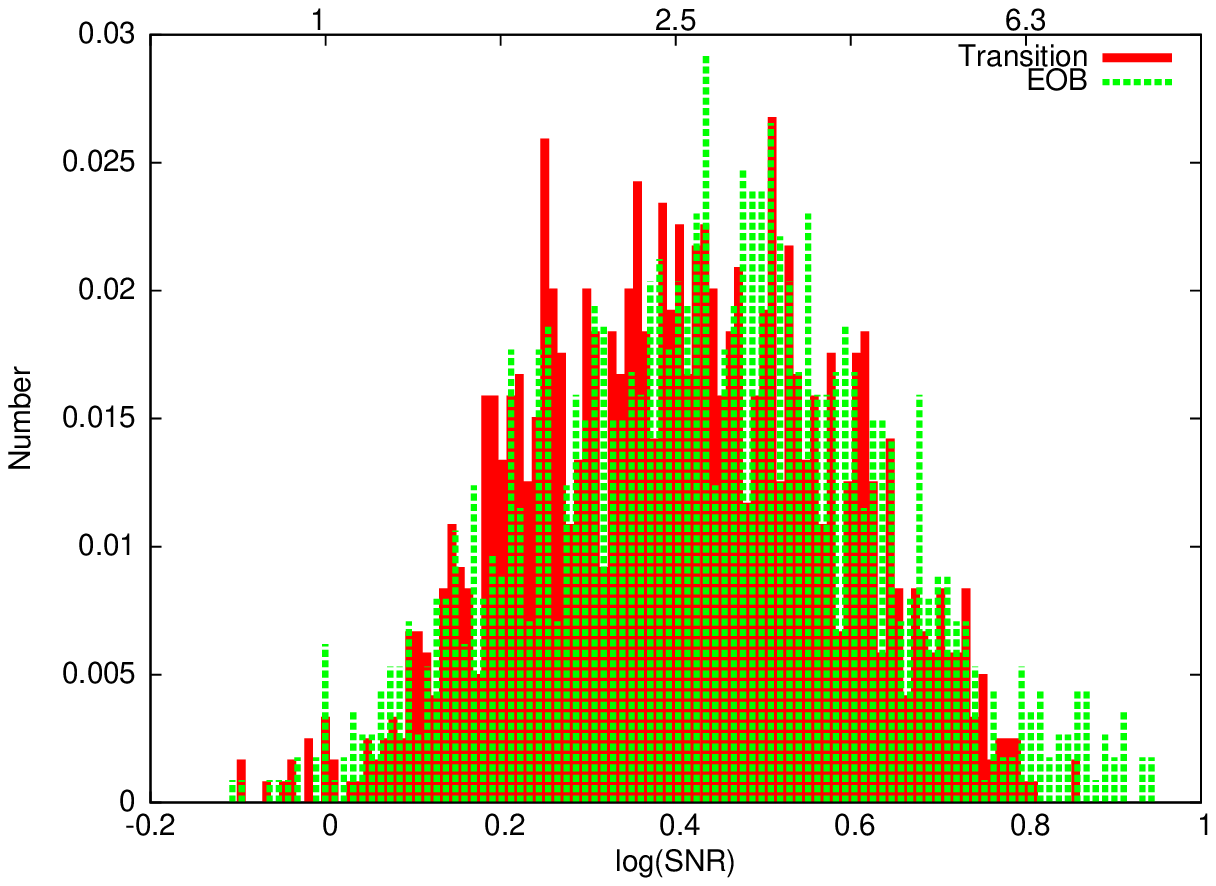}
}
\caption{ As Figure~\ref{figura3}, but now for binaries with masses \(1.4\,M_{\odot} + 500 M_{\odot}\). The left panel shows SNR distributions for IMBH spin $q=0.3$ and $q=0.9$, while the right panel shows SNR distributions for IMBH spin $q=0$ computed using both the EOB and transition models. 
}
\label{figura6}
\end{figure*}

If we consider white noise, so that \(S_h(f)\) is a constant,  we can use Parseval's theorem to rewrite the inner product \eqref{inp} as \(2 S_n^{-1}\sum_{\alpha}\int_{-\infty}^{\infty} \, p_\alpha(t) q_\alpha(t)\, {\rm d}t\) and the optimal value of the SNR becomes
\begin{equation}\label{35}
{\rm SNR}^2= 2\sum_{\alpha=I,II}\int_{t_{\rm init}}^{t_{\rm LSO}}
\hat h_{\alpha}^2(t)dt.
\end{equation}
\noindent We can use this expression for the SNR if we use noise--weighted time--domain waveforms, as defined in Eq.~\eqref{30}. We have used Eq. \eqref{35} to compute SNRs for twelve sample binary systems at a fixed distance \(D=6.63481\) Gpc, which corresponds to the luminosity distance to redshift \(z=1\). As mentioned earlier, the triangular design of ET, consisting of three interferometers with 60 degree opening angles, generates a response equivalent to two co--located interferometers with 90 degree opening angles, but rotated 45 degrees with respect to each other, and with SNR that is a factor \(3/(2\sqrt2) \sim 1.06\) higher. We have not included this factor in the quoted SNRs, given the uncertainties in the ET design that exist at this stage. The distribution of SNRs over random choices of the extrinsic parameters are summarised in Figures \ref{figura3}--\ref{figura6}, while the statistics of the SNR distributions are given in Tables~\ref{snrsI} and \ref{snrsII}.  
   
\begin{table}[thb]
\begin{tabular}{|c|c|c|c|c|c|c|c|c|}
\hline\multicolumn{1}{|c|}{}&\multicolumn{4}{c|}{$m=10M_{\odot}$}&\multicolumn{4}{c|}{$m=1.4M_{\odot}$}\\\cline{2-9}
\multicolumn{1}{|c|}{Stats}&$q=0.9$&$q=0.3$&$q=0$&$\rm EOB$&$q=0.9$&$q=0.3$&$q=0$&$\rm EOB$\\\hline
Mean       &74.645&46.774&37.844&38.282&24.099&14.388&11.272&11.967\\\cline{1-9}
St. Dev.   &31.030&20.665&12.401&14.450&9.113&5.149&4.101&4.103\\\cline{1-9}
L. Qt.     &55.081&33.266&27.606&27.669&18.113&10.740&8.337&8.933\\\cline{1-9}
U. Qt.     &103.276&65.163&52.000&51.642&32.584&19.454&15.205&16.144\\\cline{1-9}
Med.       &73.451&46.026&37.844&38.107&24.266&14.289&11.041&11.995\\\hline

\end{tabular}
\caption{Summary statistics of the SNR distributions for binary systems with a central IMBH of mass \(M=100 M_{\odot}\), and various choices for IMBH spin, $q$, and CO mass, $m$.  We show the mean, standard deviation, median and quartiles of the distribution of the SNR for each system. The SNR distributions have been computed at a fixed distance \(D=6.63481\) Gpc.  }
\label{snrsI}
\end{table}

\begin{table}[thb]
\begin{tabular}{|c|c|c|c|c|c|c|c|c|}
\hline\multicolumn{1}{|c|}{}&\multicolumn{4}{c|}{$m=10M_{\odot}$}&\multicolumn{4}{c|}{$m=1.4M_{\odot}$}\\\cline{2-9}
\multicolumn{1}{|c|}{Stats}&$q=0.9$&$q=0.3$&$q=0$&$\rm EOB$&$q=0.9$&$q=0.3$&$q=0$&$\rm EOB$\\\hline
Mean       &55.463&18.408&12.853&13.677&15.417&3.908&2.570&2.729\\\cline{1-9}
St. Dev.   &21.337&6.723&4.271&4.716&5.571&1.256&1.102&1.113\\\cline{1-9}
L.Qt.      &39.811&13.583&9.484&9.908&11.508&2.884&1.932&2.084\\\cline{1-9}
U.Qt.      &74.645&25.645&17.579&19.099&20.845&5.272&3.327&3.707\\\cline{1-9}
Med.       &55.590&18.408&12.764&13.836&15.668&3.882&2.636&2.723\\\hline

\end{tabular}
\caption{As Table~\ref{snrsI}, but now for binary systems with a central IMBH of mass \(M=500 M_{\odot}\). }
\label{snrsII}
\end{table}

\begin{figure*}[!htp]
\centerline{
\includegraphics[width=0.55\textwidth, angle=0, clip]{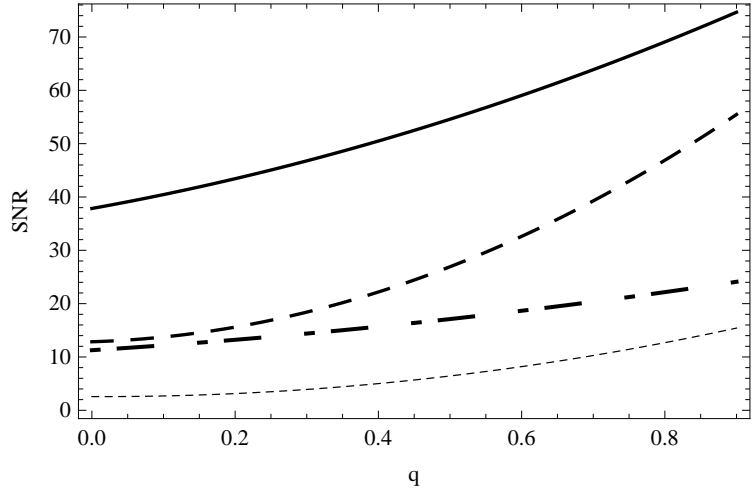}
}
\caption{The plot shows the mean value of the SNR distribution as a function of the spin parameter \(q\). From top to bottom, the various lines correspond to binaries of mass--ratio \(\eta\), namely, solid line \(\eta=0.0826\) ([10+100]\(M_{\odot}\)), dashed lined \(\eta=0.0192\) ([10+500]\(M_{\odot}\)), dash--dot line \(\eta=0.0136\) ([1.4+100]\(M_{\odot}\)) and dotted line \(\eta=0.0028\) ([1.4+500]\(M_{\odot}\)). }
\label{figurasum}
\end{figure*}

We can better visualize the results shown on Tables~\ref{snrsI} and \ref{snrsII} by plotting the mean of the SNR distributions as a function of the spin parameter \(q\), which is shown in Figure~\ref{figurasum}. We see that rapidly spinning binaries, \(q\sim 0.9\), with large mass ratios, \(\eta \sim 0.08\), will be relatively loud. At a fixed SNR detection threshold of 10, we conclude that such sources would be seen to distances  \(D\gtrsim 6.6\) Gpc. In contrast, slowly rotating binaries with \(q \sim 0.3\),  and small mass ratios, \(\eta \sim 0.003\), will only be visible at distances \(D \lesssim 6.6\) Gpc.

In order to understand the implication of these results, we can use the data summarized on Tables~\ref{snrsI} and \ref{snrsII} to estimate the number of events per year that could be detected by the ET. To do so we will follow the procedure outlined in Section 3.3 of \cite{etgair}. The basic idea is the following: fix an SNR detection threshold, $\rho_{\rm thresh}$ (we use $\rho_{\rm thresh}=10$ in the following), then estimate the luminosity distance, \(D_L(z)\), at which a given source can be detected, i.e., have this particular SNR, as $D_L = \rho(6.63481)/\rho_{\rm thresh}$, where $ \rho(6.63481)$ is the SNR of the source at a distance of $6.63481$Gpc given in the Tables. We then use the concordance cosmology to convert this luminosity distance estimate into a redshift estimate \(z\) by inverting the following expression,
\begin{equation}
\label{DLz}
D_L(z) = D_H (1+z) \left\{\int_0^z\frac{dz^\prime}{\left[\Omega_M(1+z^{\prime})^3+
\Omega_\Lambda\right]^{1/2}}\right\}.
\end{equation}
\noindent We will assume a flat universe \(\Omega_k=0\), and use $\Omega_M=0.27$, $\Omega_\Lambda=0.73$, $H_0=72$~km~s$^{-1}$~Mpc$^{-1}$; and $D_H=c/H_0\approx 4170$~Mpc. We can then compute the comoving volume, \(V_c\), within which the source can be detected using \cite{hogg},
\begin{equation}  
\label{Vc}
V_c=\frac{4\pi D_H^3}{3}\left\{\int_0^z\frac{dz^\prime}{\left[\Omega_M(1+z^{\prime})^3+
\Omega_\Lambda\right]^{1/2}}\right\}^3.
\end{equation}
In a previous study on the possible detection of IMRIs of COs into IMBHs with Advanced LIGO~\cite{manbro}, it was found that binary tightening via three--body interactions was the dominant mechanism that led to the formation of IMRIs. In this mechanism, the merger time for an IMRI can be estimated as the sum of the hardening timescale, \(T_{\rm harden}\),  and the gravitational--wave merger timescale, \(T_{\rm GW}\), i.e, $T_{\rm merge}=T_{\rm harden}+T_{\rm GW}$ \cite{etgair}, where 
\begin{eqnarray}
T_{\rm harden} &\approx& 
         2\times 10^8 \frac{10^{5.5}\ {\rm pc}^{-3}}{n} \frac{10^{13}\ {\rm cm}}{a}
        \frac{\sigma}{10\ {\rm km/s}} \frac{0.5\ M_\odot}{m_*} \ {\rm yr},
\label{Thard} \\
T_{\rm GW} &\approx& 
        10^8 \frac{M_\odot}{m} \left(\frac{100\ M_\odot}{M}\right)^2
                \left( \frac{a}{10^{13}\ {\rm cm}}\right)^4  \ {\rm yr},
\label{Tgravwave}
\end{eqnarray}
\noindent  in which \(a\) is the semimajor axis of the binary, \(n\) is the  number density of stars in a globular cluster, \(\sigma\) is the velocity dispersion, and \(m_*\) stands for the mass of stars that interact with the binary. In practice, we set $n$, $\sigma$ and $m_*$ to their fiducial values, and minimize  $T_{\rm merge}=T_{\rm harden}+T_{\rm GW}$ over $a$ to estimate the CO--IMBH coalescence rate. The rate at which IMRIs occur per globular cluster can then be approximated by $1/T_{\rm merge}$.

The SNRs that we calculate are the SNRs for systems with {\it redshifted masses}, \(M_z = M(1 + z)\), \(m_z = m(1 + z)\), equal to those that we have specified. The maximum detectable redshift estimated from the SNR then tells us the intrinsic source--frame masses of the system whose range we have computed. In the case of ET, which can detect sources out to cosmological redshifts, $z \gtrsim 1$, these intrinsic masses may not correspond to astrophysically interesting systems. However, by considering a range of redshifted masses and computing for each one an event rate under the assumption that all IMRI systems were of that particular intrinsic type, we can still obtain a rough estimate of the event rate.  

Following \cite{etgair}, we assume that \(10\%\) of clusters form an IMBH and are sufficiently dense to be hosts to an IMRI and we assume that such globular clusters have a fixed comoving density of \(\sim 0.3 {\rm  Mpc}^{-3}\). The rate of detectable events for a particular type of system can then be estimated as $\sim 0.3 (V_c/{\rm Mpc}^3) / [T_{\rm merge}(1+z)]$ \cite{etgair}. Table~\ref{evrates} presents estimates, for each of our fiducial systems, of the maximum detectable redshift, the intrinsic masses that the source represents at that redshift, the corresponding merger time through binary hardening, the comoving volume within that redshift and the IMRI event rate assuming all IMRI sources were of that type. Note that the latter assumption means that the entries in this table are not independent of each other, i.e., the total number of events is not given by the sum of the last column, but is somewhere in the range of rates tabulated. Even though the astrophysical properties of IMBHs, e.g., their mass and spin distributions, are currently very uncertain, we can still draw conservative predictions from Table~\ref{evrates}.  These estimates compare fairly well with those of \cite{etgair}, but are a bit bigger since we use a more accurate waveform model. We see that IMRIs could be seen at redshifts \(z \sim \) 1--6, depending on the mass of the IMBHs that exist, and there could be as many as a few hundred systems observed. If IMRIs tended to be $2M_\odot + 100 M_{\odot}$ systems, they could be seen out to $z \sim 4$ and we'd expect a few hundred events, but if IMRIs tended to be $1M_{\odot}+400M_\odot$ systems, we would only see $\sim10$ events out to $z \sim 0.3$. The greatest uncertainty in these figures comes from the unknown number of IMBHs that exist in the Universe, and these uncertainties are not folded into the numbers in Table~\ref{evrates}. If IMBHs are rare, then the IMRI rate could be orders of magnitude lower.  

\begin{table}[htb]
\begin{tabular}{|c|c|c|c|c|c|c|c|c|c|}
\hline
$M_z/M_\odot$ & $m_z/M_\odot$ & q &$D$/Gpc & z & $M/M_\odot$ & $m/M_\odot$ &
$T_{\rm merge}$/yr & $V_c$/Mpc$^3$ & Events/yr\\\hline
100 & 10  & 0.9 & 49.29 & 5.15 & 16.3  & 1.6 & $5.40 \times 10^8$ & $2.16 \times 10^{12}$ & 195\\\cline{1-10}
100 & 10  & 0.3 & 31.03 & 3.49 & 22.3  & 2.2 & $4.47 \times 10^8$ & $1.38 \times 10^{12}$ & 206\\\cline{1-10}
100 & 10  & 0   & 25.01 & 2.92 & 25.5  & 2.5 & $4.12 \times 10^8$ & $1.09 \times 10^{12}$ & 201\\\cline{1-10}
100 & 1.4 & 0.9 & 15.93 & 2.02 & 33.1  & 0.5 & $5.13 \times 10^8$ & $6.15 \times 10^{11}$ & 119\\\cline{1-10}
100 & 1.4 & 0.3 & 9.47  & 1.33 & 42.9  & 0.6 & $4.46 \times 10^8$ & $2.82 \times 10^{11}$ & 81\\\cline{1-10}
100 & 1.4 & 0   & 7.47  & 1.10 & 47.6  & 0.7 & $4.15 \times 10^8$ & $1.88 \times 10^{11}$ & 64\\\cline{1-10}
500 & 10  & 0.9 & 36.75 & 4.02 & 99.6  & 2.0 & $2.50 \times 10^8$ & $1.64 \times 10^{12}$ & 392\\\cline{1-10}
500 & 10  & 0.3 & 12.30 & 1.64 & 189.3 & 3.8 & $1.70 \times 10^8$ & $4.24 \times 10^{11}$ & 283\\\cline{1-10}
500 & 10  & 0   & 8.51  & 1.22 & 225.2 & 4.5 & $1.54 \times 10^8$ & $2.35 \times 10^{11}$ & 207\\\cline{1-10}
500 & 1.4 & 0.9 & 10.19 & 1.41 & 207.5 & 0.6 & $2.37 \times 10^8$ & $3.16 \times 10^{10}$ & 16\\\cline{1-10}
500 & 1.4 & 0.3 & 2.55  & 0.46 & 342.5 & 1.0 & $1.75 \times 10^8$ & $2.24 \times 10^{10}$ & 26\\\cline{1-10}
500 & 1.4 & 0   & 1.66  & 0.32 & 378.8 & 1.1 & $1.65 \times 10^8$ & $8.35 \times 10^{9}$ & 11\\\hline
\end{tabular}
\caption{``3 ET detector network'' average range, corresponding redshift, source-frame masses, merger timescale, comoving volume within range, and detectable event rate for several combinations of plausible redshifted CO and IMBH
masses.} 
\label{evrates} 
\end{table}

As discussed earlier, there is currently some uncertainty about the low-frequency sensitivity that ET will achieve. The preceding results assumed a frequency cut-off at $5$Hz, but it is informative to consider how the SNR changes if this frequency cut-off is pushed down to $3$Hz or even $1$Hz. In Table~\ref{extsnr} we show how the SNR changes as a function of the cut-off frequency, for each of the binary systems. In each case we have chosen a particular random, but fixed, set of extrinsic parameters that is representative of the events that lie around the peak of the SNR distribution for the binary.

\begin{table}[thb]
\begin{tabular}{|c|c|c|c|c|c|c|c|c|c|}
\hline\multicolumn{1}{|c|}{}&\multicolumn{3}{c|}{$q=0.9$}&\multicolumn{3}{c|}{$q=0.3$}&\multicolumn{3}{c|}{$q=0$}\\\cline{2-10}
\multicolumn{1}{|c|}{Binary}&${\rm f}= 5{\rm Hz}$&${\rm f}= 3{\rm Hz}$&${\rm f}= 1{\rm Hz}$&${\rm f}= 5{\rm Hz}$&${\rm f}= 3{\rm Hz}$&${\rm f}= 1{\rm Hz}$&${\rm f}= 5{\rm Hz}$&${\rm f}= 3{\rm Hz}$&${\rm f}= 1{\rm Hz}$\\\hline
$[10+100] M_{\odot}$  &72.111&79.799&86.896&52.240&56.885&63.680&42.855&45.186&48.306\\\cline{1-10}
$[10+500] M_{\odot}$  &55.286&58.325&63.293&19.364&20.701&22.961&15.276&15.959&16.331\\\cline{1-10}
$[1.4+100] M_{\odot}$ &20.941&21.878&23.714&16.711&17.418&18.030&11.351&11.952&12.123\\\cline{1-10}
$[1.4+500] M_{\odot}$ &13.274&13.804&13.932&4.207&4.256&4.613&2.610&2.792&3.029\\\hline
\end{tabular}
\caption{Summary of how the SNR of a source changes as the low-frequency sensitivity cut-off changes from $5$Hz to $3$Hz to $1$Hz. For each binary, the extrinsic parameters have been chosen randomly, but kept fixed as the cut-off frequency was varied.}
\label{extsnr}
\end{table}

\begin{figure*}[!htp]
\centerline{
\includegraphics[width=0.55\textwidth, angle=0, clip]{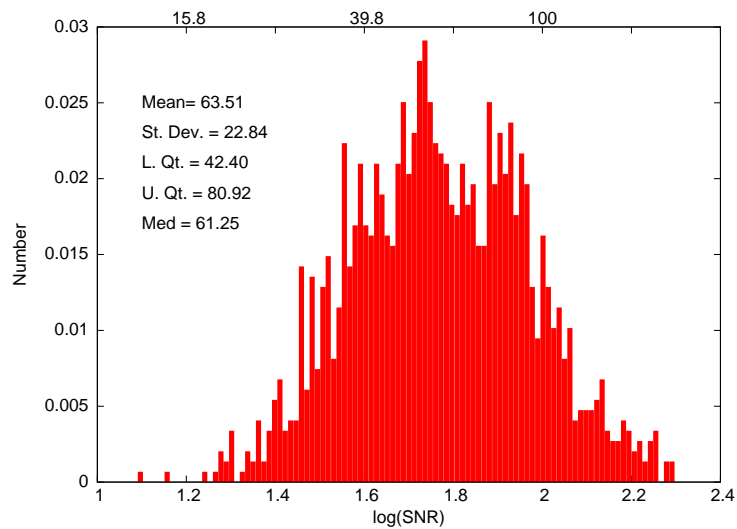}
}
\caption{ The panel shows the SNR distribution  for a \(10\,M_{\odot}+100M_{\odot}\) binary with IMBH spin parameter \(q= 0.3\), at a frequency cut--off of 1Hz.  The horizontal axis is the logarithm to base ten of the SNR. Note that the SNR distribution has been computed at a fixed distance \(D=6.63481\) Gpc, or, equivalently, at a fixed redshift \(z=1\). The top horizontal axis mirrors the values used in the bottom horizontal one, but in normal units. The statistics of the system are summarized at the top--left of the panel.}
\label{ccheck}
\end{figure*}

We see that if ET achieves frequency sensitivity down to 1Hz, we will have somewhat greater sensitivity to systems containing more massive compact objects. This will facilitate the extraction of these signals from the data, and open up the possibility of detecting these sources at higher redshift. We have checked the reliability of the figures quoted in Table~\ref{extsnr} by running a full Monte Carlo for the particular system $[10+100] M_{\odot}$, with spin parameter \(q=0.3\), and frequency cut--off at 1Hz. Figure~\ref{ccheck} presents these results, and confirms that Table~\ref{extsnr} is a fair representation of what may be achieved by ET at lower frequencies.

We can use the figures quoted in Table~\ref{extsnr} to explore the redshift at which the loudest sources could be seen if ET achieved a frequency cut--off of 1Hz. From Table~\ref{evrates}, we already know that the systems with redshifted masses, \(m_z + M_z\), of $[10+100] M_{\odot}$, $[10+500] M_{\odot}$, with \(q=0.9\), could be seen up to redshift  \(z\sim 5\), \(z\sim 4\), respectively. Using Eq.~\eqref{DLz}, we see that, with a cut-off at 1Hz, these same systems could be detected up to redshift \(z\sim 6\), \(z\sim 5\), respectively. Table~\ref{1hzq09} summarizes these results for the loudest sources from Table~\ref{evrates}. The event rate estimate only changes appreciably for the $q=0.3$ case (since the intrinsic masses for the systems are also changing), but as $q=0.3$ may be a good estimate for the typical spin of an IMRI, there could be a significant scientific gain from pushing the lower frequency cut-off to $1$Hz. A more systematic study, which fixes the intrinsic masses of the events as opposed to the redshifted masses, is needed to fully explore the implications of the cut-off on the expected IMRI detection rate.

\begin{table}[htb]
\begin{tabular}{|c|c|c|c|c|c|c|c|c|c|}
\hline
$M_z/M_\odot$ & $m_z/M_\odot$ & q &$D$/Gpc & z & $M/M_\odot$ & $m/M_\odot$ &
$T_{\rm merge}$/yr & $V_c$/Mpc$^3$ & Events/yr\\\hline
100 & 10  & 0.9 & 57.72 & 5.9 & 14.5  & 1.4 & $5.81 \times 10^8$ & $2.47 \times 10^{12}$ & 187\\\cline{1-10}
100 & 10  & 0.3 & 42.46 & 4.6 & 17.8  & 1.8 & $4.47 \times 10^8$ & $1.92 \times 10^{12}$ & 202\\\cline{1-10}
500 & 10  & 0.9 & 41.79 & 4.5 & 90.9  & 1.8 & $2.62 \times 10^8$ & $1.87 \times 10^{12}$ & 385\\\cline{1-10}
500 & 10  & 0.3 & 15.26 & 2.2 & 156.2 & 3.1 & $1.70 \times 10^8$ & $7.08 \times 10^{11}$ & 346\\\hline

\end{tabular}
\caption{``3 ET detector network'' average range, corresponding redshift, source-frame masses, merger timescale, comoving volume within range, and detectable event rate for several combinations of plausible redshifted CO and IMBH masses. The cutt--off frequency has been set at 1Hz.} 
\label{1hzq09} 
\end{table}

There is another important consequence of a lower low-frequency cut-off. As discussed previously, the binary systems we have considered in this paper are very short lived. However, by pushing the seismic wall down to 1Hz these binaries now stay in the sensitivity band of the detector for longer. Therefore, we should be able to obtain better parameter estimation accuracies, in particular for the extrinsic parameters. Table~\ref{exttime} gives the time that each of these binaries is in band as a function of the frequency cut--off.

\begin{table}[thb]
\begin{tabular}{|c|c|c|c|c|c|c|c|c|c|}
\hline\multicolumn{1}{|c|}{}&\multicolumn{3}{c|}{$q=0.9$}&\multicolumn{3}{c|}{$q=0.3$}&\multicolumn{3}{c|}{$q=0$}\\\cline{2-10}
\multicolumn{1}{|c|}{Binary}&${\rm f}= 5{\rm Hz}$&${\rm f}= 3{\rm Hz}$&${\rm f}= 1{\rm Hz}$&${\rm f}= 5{\rm Hz}$&${\rm f}= 3{\rm Hz}$&${\rm f}= 1{\rm Hz}$&${\rm f}= 5{\rm Hz}$&${\rm f}= 3{\rm Hz}$&${\rm f}= 1{\rm Hz}$\\\hline
$[10+100] M_{\odot}$  &45.0&169.2&3093.7&42.1&161.5&3048.5&38.9&157.3&3025.8\\\cline{1-10}
$[10+500] M_{\odot}$  &16.3&61.2&1099.3&8.3&43.9&1010.1&5.4&35.6&963.2\\\cline{1-10}
$[1.4+100] M_{\odot}$ &319.9&1209.0&22089.1&291.0&1152.4&21776.9&275.9&1123.7&21614.1\\\cline{1-10}
$[1.4+500] M_{\odot}$  &112.5&436.1&7851.8&55.6&311.2&7221.9&34.0&252.5&6879.9\\\hline
\end{tabular}
\caption{Summary of how the time spent in band of a source changes as the low-frequency sensitivity cut-off changes from $5$Hz to $3$Hz to $1$Hz.   The time is reported in seconds. For each binary, the extrinsic parameters have been chosen randomly, but kept fixed as the cut-off frequency was varied.}
\label{exttime}
\end{table}

Table~\ref{exttime} shows that by pushing the frequency cut--off to 1Hz will boost the time spent in band of the shortest--lived events  by a factor of \(\sim 200 \). Binaries of $[1.4+100] M_{\odot}$ could spend up to 6 hours in band. In the second paper of this series we will show that even with a 5Hz cut-off we can pinpoint the location of these particular  sources in the sky and constrain their luminosity distances with a precision of \(\sim10\%\), at SNR of 30, using a 3 ET detector network. Pushing the seismic limit down to 1Hz will further improve the parameter estimation accuracies. 

All the results obtained up to this point have assumed that we will be able to detect IMRIs using a 3 ET detector network. Hence, our previous results may be considered as upper limits for the various quantities we have quoted --- event rates per year, SNRs, horizon distances etc. We shall now relax this assumption and explore more modest scenarios, the configurations C1--C5 outlined earlier. To recap, these are, C1: one ET at the geographic location of Virgo; C2: as configuration C1 plus a right--angle detector at the location of LIGO Livingston;  C3: as configuration C1 plus another ET at the location of LIGO Livingston;  C4: as configuration C2 plus another right--angle detector in Perth; and C5: as configuration C3 plus another ET in Perth. Note that configuration C5 corresponds to the 3ET detector network used in our analyses so far. We will quote results for IMBHs with spin parameter \(q\sim0.3\) only, since as argued earlier this could be a reasonable fiducial value for IMBH spin. We quote SNRs for IMRIs into a central black hole of mass $M=100M_{\odot}$ and spin $q=0.3$ for five different configurations in Table~\ref{newsnrsI} and corresponding results for IMRIs with $M=500M_{\odot}$ and $q=0.3$ in Table~\ref{newsnrsII}.

\begin{table}[thb]
\begin{tabular}{|c|c|c|c|c|c|c|c|c|c|c|}
\hline\multicolumn{1}{|c|}{}&\multicolumn{5}{c|}{$m=10M_{\odot} $}&\multicolumn{5}{c|}{$m= 1.4M_{\odot} $}\\\cline{2-11}
\multicolumn{1}{|c|}{Stats}&${\rm C1}$&${\rm C2}$&${\rm C3}$&${\rm C4}$&${\rm C5}$&${\rm C1}$&${\rm C2}$&${\rm C3}$&${\rm C4}$&${\rm C5}$\\\hline
Mean       &26.242&34.119&40.179&40.458&46.774&8.222&10.740&12.589&12.823&14.388\\\cline{1-11}
St. Dev.   &14.785&16.623&18.912&19.015&20.665&3.113&3.485&4.102&4.408&5.149\\\cline{1-11}
L.Qt.      &18.239&24.660&28.708&29.242&33.266&5.754&8.054&9.333&9.311&10.740\\\cline{1-11}
U. Qt.     &38.282&48.195&57.148&55.719&65.163&11.169&14.928&16.711&16.634&19.454\\\cline{1-11}
Med.       &26.424&33.884&39.719&40.365&46.026&8.017&10.666&12.162&12.445&14.289\\\hline
\end{tabular}
\caption{As Table~\ref{snrsI}, but for binary systems with a central IMBH of mass \(M=100 M_{\odot}\) and spin parameter \(q=0.3\) and assuming four additional configurations for the detector network, C1--C5 as described in Section~\ref{ETdes}. Configuration C5 is the network of three ETs which has been used for all results elsewhere in this paper.}
\label{newsnrsI}
\end{table}

\begin{table}[thb]
\begin{tabular}{|c|c|c|c|c|c|c|c|c|c|c|}
\hline\multicolumn{1}{|c|}{}&\multicolumn{5}{c|}{$m= 10M_{\odot}$}&\multicolumn{5}{c|}{$m=1.4M_{\odot} $}\\\cline{2-11}
\multicolumn{1}{|c|}{Stats}&${\rm C1}$&${\rm C2}$&${\rm C3}$&${\rm C4}$&${\rm C5}$&${\rm C1}$&${\rm C2}$&${\rm C3}$&${\rm C4}$&${\rm C5}$\\\hline
Mean       &9.528&12.677&14.825&14.894&18.408&2.084&2.685&3.141&3.177&3.908\\\cline{1-11}
St.Dev.    &4.432&4.866&5.551&5.346&6.723&1.102&1.117&1.164&1.129&1.256\\\cline{1-11}
L.Qt.      &6.683&8.974&10.328&12.677&13.583&1.517&2.032&2.333&2.382&2.884\\\cline{1-11}
U.Qt.      &13.868&17.742&20.370&20.045&25.645&2.844&3.589&4.188&4.207&5.272\\\cline{1-11}
Med.       &9.638&12.853&14.521&14.689&18.408&2.109&2.704&3.090&3.126&3.882\\\hline
\end{tabular}
\caption{As Table~\ref{newsnrsI}, but for binary systems with a central IMBH of mass \(M=500 M_{\odot}\) and spin parameter \(q=0.3\).}
\label{newsnrsII}
\end{table}

From Tables~\ref{newsnrsI} and \ref{newsnrsII} we learn that the SNR corresponding to a 3 ET detector network is, roughly speaking, a factor of \(\sqrt3\) and  \(\sqrt{3/2}\) greater than the SNR associated with a single ET and a 2 ET network, respectively. We may expect this scaling since, despite the different locations of the detectors, we would expect the SNR to scale approximately as the square root of the number of detectors. We also notice that the SNR associated with configuration C4 is similar to the SNR of the 3 ET network. This configuration is somewhat less ambitious than a 3ET network and hence might be more likely to be realised in the future. We will show in the second article of this series that for slowly rotating IMBHs, such a network will also suffice for determining the luminosity distance to an event with an accuracy of \(\sim 12\%\) at a source SNR of 30.  If the luminosity distance is converted into a redshift using the concordance cosmology at that time, then the resulting redshift error will be comparable to the distance error. Furthermore, configuration  C4 will also allow us to determine the masses of the components of the system to high accuracy. 

The above discussion indicates that ET, through the detection of gravitational waves emitted by IMRIs of COs into IMBHs, may shed some light on the astrophysical properties of IMBHs, e.g., their mass and spin distributions, and to find out whether dynamical interactions in dense stellar systems do indeed lead to the formation of IMBHs. Additionally, the  mergers detected by ET will be complementary to mergers between heavier BHs that will be seen by space--based detectors such as LISA, ALIA or DECIGO. These various detectors will provide a measurement of the rate of black hole mergers in various mass ranges, which will be useful  to place constraints on models of black hole growth~\cite{etgair}.

\section{Conclusions}
\label{s6}
We have developed waveform models for IMRIs of COs on circular, equatorial orbits into central IMBHs of arbitrary spin. One waveform model, which is valid for inspirals into IMBHs of arbitrary spin, uses the transition-to-plunge scheme of Ori and Thorne~\cite{amos} to smoothly match the inspiral onto a plunge waveform and ringdown. The second approach, at present valid only for non-spinning IMBHs, uses the effective-one-body formalism to match a merger and ringdown onto the inspiral waveform. We have shown that the two distinct waveform families are in good agreement in the \(q=0\) limit, particularly in the small  \(\eta\) regime. The agreement in phasing all the way from inspiral to ringdown is particularly good and this is important as a good phase model will be crucial for the detection of these systems via matched filtering and to extract parameter information from the detector measurements. 

We have used these waveform models to compute estimates for the SNR that would be obtained in ET for various binary systems. We computed SNRs for a 3 ET detector network for twelve different binaries using the transition model and used the EOB model to cross--check the results for the non--spinning IMBHs. We found that the two models made predictions that were consistent to about ten percent. Assuming that ET has a low-frequency sensitivity cut--off at 5Hz, we found that at a redshift of $z=1$, typical SNRs for IMRI systems with masses $1.4M_\odot$+$100M_{\odot}$, $10M_\odot$+$100M_{\odot}$, $1.4M_\odot$+$500M_{\odot}$ and $10M_\odot$+$500M_{\odot}$ will be in the range $\sim 10$--$25$, $\sim40$--$80$, $\sim3$--$15$ and $\sim 10$--$60$ respectively. Using the SNR distributions as input data, we estimated the horizon distance at which these various sources could be seen. This suggested that ET could detect as many as several hundred of these systems, up to a redshift  \(z\lesssim 5\), although the exact number will depend on the intrinsic distribution of masses and spins for the IMRI systems. If the ET sensitivity extends down to 1Hz, we found that the same systems could be detected up to redshift \(z\lesssim 6\).

We have also explored more modest network configurations consisting of 1 ET only, 1 ET plus 1 right--angle detector, 2 ETs and 1 ET plus 2 right--angle detectors. Using these configurations we computed SNR distributions for the same four combinations of source masses  $1.4M_\odot$+$100M_{\odot}$, $10M_\odot$+$100M_{\odot}$,  $1.4M_\odot$+$500M_{\odot}$, $10M_\odot$+$500M_{\odot}$, but fixed IMBH spin parameter \(q=0.3\). We chose this spin since it is a reasonable estimate of the spin parameter of an IMBH that has grown primarily through a series of minor mergers. A network consisting of one ET and two right-angle interferometers will have almost as great a sensitivity to IMRIs as the highly-ambitious 3-ET network, and this result will be important when a third-generation detector network is planned.

We shall see in the second paper of this series that a configuration consisting of  1 ET only will not be enough to effectively constrain the distance to the source, let alone to pinpoint the location of the source in the sky. However, the configuration consisting of 1ET and 2 right--angle detectors could determine the luminosity distance of a source to an accuracy of \(\sim15\%\) at SNR of 30. Since ET will also be able to determine the masses of a binary to great accuracy, we should be able to estimate with confidence the masses and the redshift at which two COs merge. This information will be important for understanding the astrophysical properties and history of IMBHs. IMRI detections will probe the existence and properties of IMBHs, and their number density over cosmic history. This in turn will tell us how these objects form and evolve. Since IMRIs will primarily be observed from globular clusters, IMRI observations will tell us about the efficiency of formation of IMBHs in cluster environments and the number density of cluster IMBHs~\cite{etgair}. This in turn will allow us to estimate the rate at which IMBH-IMBH binaries could form through the globular cluster channel. ET might also detect IMBH-IMBH binaries that arise from primordial IMBHs~\cite{sesa}. Understanding the properties of globular cluster IMBHs through the IMRI channel could therefore help to identify candidate primordial IMBHs, which will have important consequences for hierarchical models of structure growth. IMRIs could also be used to test gravitational physics in the strong field and map the structure of spacetime outside IMBHs~\cite{etgair}. The results in this paper indicate that ET might detect as many as several hundred IMRI events which could be used to extract this physics. These results also lay a solid foundation for  paper II in this series, which will study the precision with which the parameters of these IMRI systems might be estimated using the ET detector.

\section*{Acknowledgments}
EH is funded by CONACyT. JG's work is supported by the Royal Society. EH also thanks B. Sathyaprakash and Alberto Vecchio for useful comments. 

\bibliography{allpap}

\end{document}